\documentclass[12pt, onecolumn, journal]{IEEEtran}

\usepackage{amsmath,amssymb,amscd}
\usepackage{graphicx,cite}
\usepackage{algorithm}
\usepackage{algpseudocode}
\usepackage{subfigure}
\usepackage{multirow}

\newcommand{\opt}{\text{opt}}

\begin{document}
\title{Optimization of Massive Full-Dimensional MIMO for Positioning and Communication}
\author{\IEEEauthorblockN{Seongah Jeong, Osvaldo Simeone, and Joonhyuk Kang}\\
\thanks{The work of O. Simeone was partially supported by the U.S. NSF through grant CCF-1525629. O. Simeone has also received
funding from the European Research Council (ERC) under the European Unions Horizon 2020 research and innovation programme (grant
agreement No 725731). This work was supported by the National Research Foundation of Korea (NRF) grant funded by the Korea government(MSIP) (No. 2017R1A2B2012698).

Seongah Jeong is with the School of Engineering and Applied Sciences (SEAS), Harvard University, Cambridge, MA 02138, USA (Email:
seongah116@gmail.com). 

Osvaldo Simeone is with the Department of Informatics, King’s College London, London, UK (Email: osvaldo.simeone@kcl.ac.uk). 

Joonhyuk Kang is with the Department of Electrical Engineering, Korea Advanced Institute of Science and Technology (KAIST) Daejeon, South Korea (Email: jhkang@ee.kaist.ac.kr).}
}
\maketitle
\vspace{-1.5cm}
\begin{abstract}
Massive Full-Dimensional multiple-input multiple-output (FD-MIMO) base stations (BSs) have the potential to bring multiplexing and coverage gains by means of three-dimensional (3D) beamforming. Key technical challenges for their deployment include the presence of limited-resolution front ends and the acquisition of channel state information (CSI) at the BSs. This paper investigates the use of FD-MIMO BSs to provide simultaneously high-rate data communication and mobile 3D positioning in the downlink. The analysis concentrates on the problem of beamforming design by accounting for imperfect CSI acquisition via Time Division Duplex (TDD)-based training and for the finite resolution of analog-to-digital converter (ADC) and digital-to-analog converter (DAC) at the BSs. Both \textit{unstructured beamforming} and a low-complexity \textit{Kronecker beamforming} solution are considered, where for the latter the beamforming vectors are decomposed into separate azimuth and elevation components. The proposed algorithmic solutions are based on Bussgang theorem, rank-relaxation and successive convex approximation (SCA) methods. Comprehensive numerical results demonstrate that the proposed schemes can effectively cater to both data communication and positioning services, providing only minor performance degradations as compared to the more conventional cases in which either function is implemented. Moreover, the proposed low-complexity Kronecker beamforming solutions are seen to guarantee a limited performance loss in the presence of a large number of BS antennas.      
\end{abstract}

\section{Introduction}\label{sec:intro}
Mobile broadband communication in the New Radio (NR) physical layer of 5G systems is not only expected to increase data rate and reliability, but also to cater to new services, including proactive radio resource management, intelligent traffic systems, autonomous vehicles, Internet of Things (IoT), and device-to-device communication for disaster response and emergency relief. Services such as these can benefit from location awareness at the mobile users \cite{Taranto14SPMAG, Lin17Arxiv}. According to \cite{5GPPP}, 5G is envisioned to attain positioning accuracy of one meter or less, outperforming existing positioning techniques such as GPS and wireless local area network (WLAN) fingerprinting-based systems. 

The support for positioning in Long-Term Evolution (LTE) systems has been standardized in the form of downlink observed time difference of arrival (OTDOA) in Release 9 and unlink TDOA (UTDOA) in Release 11, with additional work on dedicated signals, procedures, and requirements for vertical localization accuracy to be carried out for in Release 14, which  marks the start of 5G \cite{Hoy16Commmag}. 

Among the key technologies introduced to boost the spectral efficiency of 5G, the use of very large antenna arrays, or massive multiple-input multiple-output (MIMO) systems, \cite{3GPP36897, Nam13COMMMAG, Li16Arxiv, Jacobsson16Arxiv, Mondal15commag, Kammoun14JSAC, Gcong16TVT, Ying14ICC} at the base stations (BSs) is of notable importance. Technical issues that challenge the deployment of massive MIMO systems include the large space occupation, the hardware cost associated with radio frequency (RF) elements, and the power dissipation of a large antenna array. As a specific solution, three-dimensional (3D) MIMO, or Full-Dimensional MIMO (FD-MIMO), has been intensely discussed in the LTE Release 13 \cite{3GPP36897, Nam13COMMMAG, Mondal15commag, Kammoun14JSAC}. FD-MIMO BSs are equipped with two-dimensional (2D) antenna arrays, thereby reducing the spatial size of the BS and providing the additional degree of freedom for beamforming design, given by the elevation angle. In addition, the problem of cost and circuit power dissipation is typically addressed by using low-resolution Analog-to-Digital converter/Digital-to-Analog converter (ADC/DAC) \cite{Li16Arxiv, Jacobsson16Arxiv} or by developing hybrid analog-digital transceiver \cite{Tsinos16asilomar}.             

In this paper, as illustrated in Fig. \ref{fig:sys}, we consider a cellular system with FD-MIMO BSs having per-antenna limited-resolution front ends. In the system, as per the frame structure in Fig. \ref{fig:ch}, each MS estimates its 3D position, as well as the downlink channels, based on the pilot signals received during the downlink training phase. It then decodes the data received from all the BSs during downlink data phase. It is noted that downlink localization may be advantageous with respect to uplink localization due to the large transmission power of the BSs \cite{WangWiMAX}. We focus on the problem of downlink beamforming design in the presence of imperfect channel state information (CSI), which is estimated at the BSs via uplink training using time-division duplex (TDD). Unlike the prior work \cite{3GPP36897, Nam13COMMMAG}, transmit downlink beamforming is optimized so as to serve both data communication and localization services. The design accounts for impairments in CSI acquisition accuracy and downlink transmission caused by low-resolution ADC/DACs at the BSs. Furthermore, we consider both general \textit{unstructured beamforming} and a low-complexity \textit{Kronecker beamforming} solution. In the latter case, the beamforming vectors are decomposed into the beams in azimuth and elevation and can be optimized separately \cite{Ying14ICC}. The proposed schemes leverage the Bussgang theorem \cite{Bussgang52} to model the effects of ADC/DAC quantization. In both cases, we consider two complementary formulations: ($i$) sum-rate maximization under localization accuracy and total transmit power constraints; or ($ii$) sum-localization error minimization under data rate and total transmit power constraints. We finally note that our prior work \cite{JSA14TVT} also tackles the problem of beamforming design for localization and data transmission, but it considers 2D localization and infinite resolution front-ends, and it concentrates solely on the problem of power minimization.   

The rest of the paper is organized as follows. We describe the system model in Section II and the performance metrics in Section \ref{sec:metric}. Then, we propose unstructured beamforming and Kronecker beamforming designs in Section \ref{sec:beam_ideal} and Section \ref{sec:beam_kron}, respectively. In Section \ref{sec:num}, numerical results are presented, and concluding remarks are summarized in Section \ref{sec:con}.   
\section{System Model}\label{sec:sys}
\begin{figure}[t]
\begin{center}
\includegraphics[width=12cm]{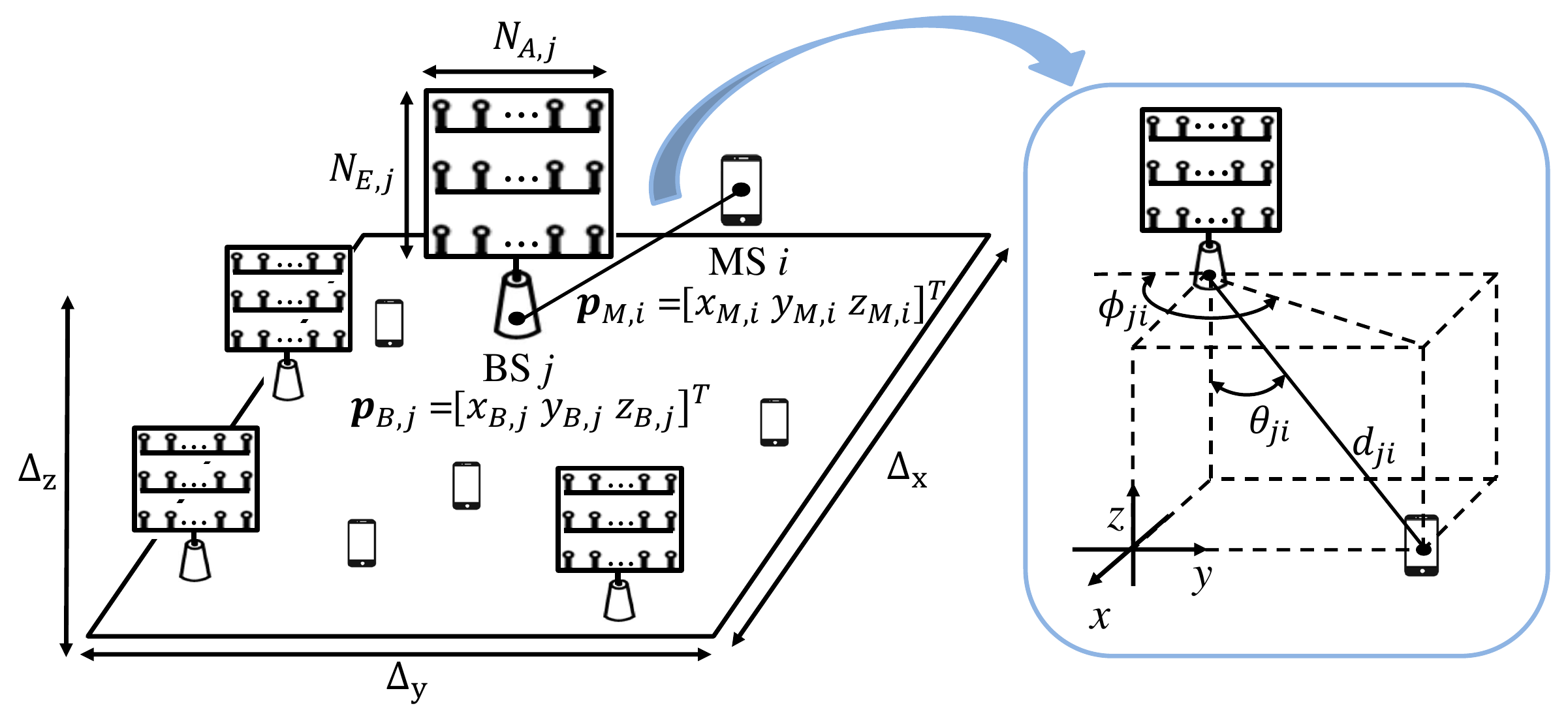}
\caption{In the system under study, $N_B$ BSs equipped with FD-MIMO arrays and a limited-resolution front end serve $N_M$ single-antenna MSs. Each MS estimates its 3D position $\pmb{p}_{M,i}$ based on the pilot signals received during the downlink training phase and receives data from all the BSs during the downlink data phase. The system operates via TDD and uplink and downlink channels are estimated at the BSs during the uplink training phase.} \label{fig:sys}
\end{center}
\end{figure}

Consider the cellular system as illustrated in Fig. \ref{fig:sys}, in which $N_B$ BSs, equipped with massive FD-MIMO antenna arrays and a limited-resolution front-end, serve $N_M$ MSs for the purpose of downlink communication and positioning. As detailed below, the system operates via TDD for the purpose of uplink-based channel estimation at the BSs \cite{Marzetta06Asilomar}, and it uses a frequency reuse scheme that assigns a different band to each BS in the area under study (see, e.g,. \cite{Erkip17Arxiv}). Our focus is the design of downlink beamforming vectors based on estimated CSI with the goal of ensuring performance guarantees in terms of both data transmission and localization in the presence of finite-resolution front-ends at the BSs.   

The set of BSs and MSs are denoted as $\mathcal{N}_B=\{1, \dots, N_B\}$ and $\mathcal{N}_M=\{1, \dots, N_M\}$, respectively. The FD, or 2D, uniform rectangular array (URA) at each BS $j \in \mathcal{N}_B$ has $N_j = N_{A,j}N_{E,j}$ antennas, where the $N_{A,j}$ horizontal antennas, placed along the $y$ axis, have spacing $d_A$, and the $N_{E,j}$ vertical antennas, located along the $z$ axis, have spacing $d_E$. Note that, as in \cite{Ying14ICC, Gcong16TVT}, we assume no mechanical downtilt for the antenna array. The MSs have a single antenna. 

We assume a digital massive MIMO implementation in which a low-resolution ADC/DAC is available for each antenna element at the BSs \cite{Mezghani12ISIT, Stein13ICASSP, Li16Arxiv, Kakkavas16WSA, Jacobsson16Arxiv}. Each ADC/DAC at BS $j$ has $B_j$ quantization levels $\mathcal{B}_j = \{ b_{j, 0}, \dots, b_{j, B_j-1}\}$ for both the in-phase and quadrature components. We define the corresponding quantization function operating separately on the in-phase and quadrature components of each element of the argument vector as $\mathcal{Q}_{B_j}(\cdot)$. 

The MS $i \in \mathcal{N}_M$ is located at position $\pmb{p}_{M,i}=[x_{M,i} \,\,y_{M,i}\,\,z_{M,i}]^T$, which is randomly and uniformly distributed within a $\Delta_x \times \Delta_y \times \Delta_z$ cube. Instead, BS $j \in \mathcal{N}_B$ is located at a fixed position $\pmb{p}_{B,j}=[x_{B,j} \,\,y_{B,j}\,\,z_{B,j}]^T$ within the cube. The positions $\{\pmb{p}_{B,j}\}_{j \in \mathcal{N}_B}$ of the BSs are assumed to be known to all the nodes in the network. The distance $d_{ji}$, azimuth angle $\phi_{ji}$ and elevation angle $\theta_{ji}$ between BS $j$ and MS $i$ are denoted as $d_{ji} = \| \pmb{p}_{M,i} - \pmb{p}_{B,j}\|$, $\phi_{ji} = \tan^{-1}\left(- \frac{x_{M,i}-x_{B,j}}{y_{M,i}-y_{B,j}}\right)$, and $\theta_{ji} = \tan^{-1}\left( - \frac{\sqrt{\left(x_{M,i}-x_{B,j}\right)^2+\left(y_{M,i}-y_{B,j}\right)^2}}{z_{M,i}-z_{B,j}}\right)$, respectively, where the angles $\phi_{ji}$ and $\theta_{ji}$ are defined with respect to the negative $y$-axis and $z$-axis. 

\begin{figure}[t]
\begin{center}
\includegraphics[width=10cm]{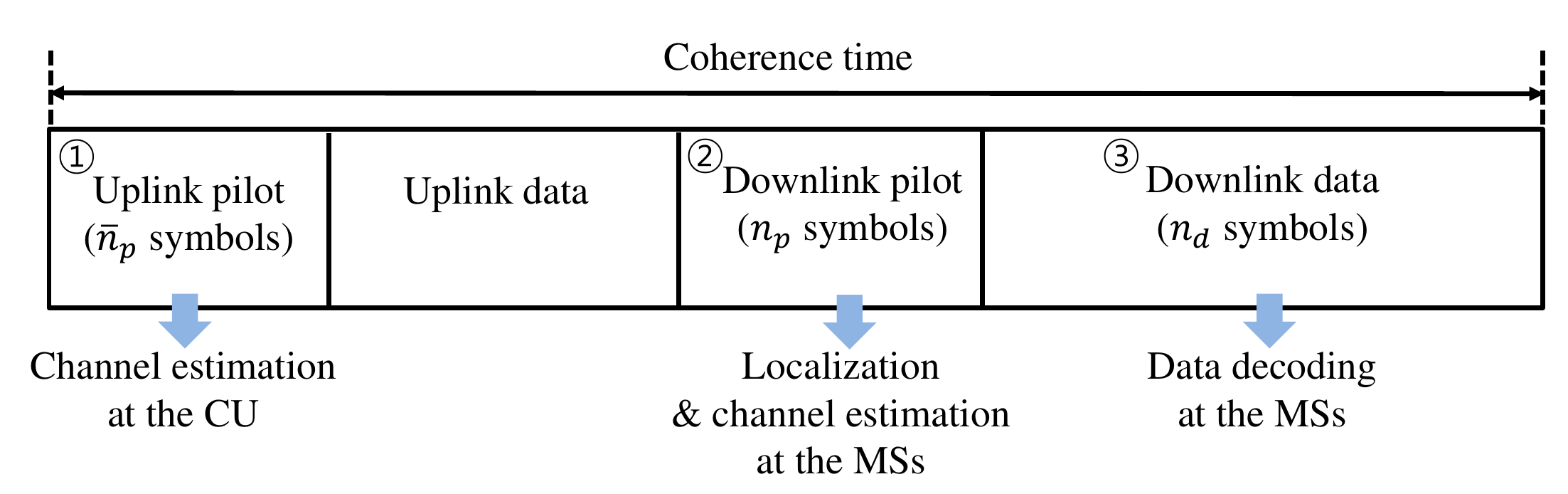}
\caption{Illustration of the frame structure in each frequency band.} \label{fig:ch}
\end{center}
\end{figure}

As in \cite{JSA14TVT}, we assume that each BS $j$ communicates with the MSs simultaneously over a frequency band $j$ that is orthogonal to the bands $i$ assigned to any other BS $i \neq j$. Note that this precludes the use of cooperative processing across BSs, such as cooperative multi-point (CoMP) transmission \cite{Erkip17Arxiv}. As shown in Fig. \ref{fig:ch}, transmission in each frequency band is organized in frames. Uplink and downlink channels in all bands are constant within each frame and change according to stationary independent ergodic processes from one frame to the next. Due to channel reciprocity, the uplink channel matrix is assumed to be equal to the transpose of the downlink channel matrix \cite{Marzetta06Asilomar}.  

Frames are divided into the following slots. $1$) \textit{Uplink training}: The MSs transmit simultaneously orthogonal pilot signals in all the bands. Each BS estimates the channels of all the MSs in the assigned band. This pilot slot, of duration $\bar{n}_p$ symbols, is followed by an uplink data slot, which is not of interest in this work. $2$) \textit{Downlink training}: Each MS uses the signal received in the downlink pilot slot to perform CSI estimation and localization. To enable 3D localization, we assume the condition $N_B \ge 4$, so that each MS $i$ can locate the position $\pmb{p}_{M,i}$ via multiangulation or multilateration based on the time measurements. Furthermore, the BSs and MSs are assumed to have a common time reference. $3$) \textit{Downlink data}: The MSs receive data in the downlink data slot from all the BSs. 

 In the following, we will use subscripts or superscripts $p$ and $d$ for variables related to pilots and data, respectively, while the bar notation, plain letters and hat notation are used for variables pertaining to the uplink transmission, downlink transmission and estimation, respectively.
\subsection{FD Channel Models}\label{sec:ch}
We will consider two types of FD channel models that are typically adopted for 3D modeling, namely the \textit{correlated channel model} \cite{Mondal15commag, Kammoun14JSAC} and the \textit{Kronecker channel model} \cite{Ying14ICC, Gcong16TVT}. 

\subsubsection{Correlated Channel Model}\label{sec:chg}
The channel between BS $j$ and MS $i$ in any given coherence time is denoted as 
\vspace{-0.5cm}
\begin{equation}\label{eq:ch}
\pmb{g}_{ji}=\zeta_{ji}\pmb{h}_{ji},
\end{equation}
where $\zeta_{ji}$ accounts for the path loss between BS $j$ and MS $i$, which can be written as 
\begin{equation}\label{eq:pl}
\zeta_{ji}=\left(1+\left(\frac{d_{ji}}{d_0}\right)^\eta\right)^{-\frac{1}{2}},
\end{equation}     
with $\eta$ being the path loss exponent and $d_0$ being a reference distance (see, e.g., \cite{Molit05Book, JSA14TVT, Gcong16TVT}). The pathloss (\ref{eq:pl}) is random, given that the MSs are spatially and randomly distributed within the given cube volume, with first and second moments $\mathbb{E}[\zeta_{ji}]=\mu_{\zeta_{j}}$ and $\mathbb{E}[\zeta_{ji}^2]=\sigma_{\zeta_{j}}^2$, respectively. Each vector $\pmb{h}_{ji} \sim \mathcal{CN}(\pmb{0}_{N_j \times 1}, \pmb{R}_{h_{ji}})$ represents correlated small-scale Rayleigh fading and is characterized by a covariance matrix $\pmb{R}_{h_{ji}} \triangleq \mathbb{E}[\pmb{h}_{ji}\pmb{h}^H_{ji}]$. The path loss and small-scale fading parameters are assumed to be independent, which yields the moments $\mathbb{E}[\pmb{g}_{ji}] = \pmb{0}_{N_j \times 1}$ and $\mathbb{E}[\pmb{g}_{ji}\pmb{g}^H_{ji}] = \sigma_{\zeta_{j}}^2\pmb{R}_{h_{ji}} \triangleq \pmb{R}_{g_{ji}}$. Following \cite{Mondal15commag, Kammoun14JSAC, Ying14ICC, Gcong16TVT}, the correlation between the $(k,l)$th antenna element and $(p,q)$th antenna element, with the $(k,l)$th antenna element indicating the $k$th in elevation and $l$th in azimuth element of the URA, is given as
\begin{equation}\label{eq:ch_R}
\left[\pmb{R}_{h_{ji}}\right]_{(k,l), (p,q)} = \frac{\gamma_1}{\sqrt{\gamma_5}}e^{-\frac{\gamma_7}{2\gamma_5}}e^{j\frac{\gamma_2\gamma_6}{\gamma_5}}e^{-\frac{(\gamma_2\sigma_{\phi}\sin\phi_{ji})^2}{2\gamma_5}},
\end{equation}  
where $\{\gamma_k\}_{k \in \{1, \dots, 7\}}$ are functions of $(k, l, p, q)$ and are calculated as in \cite{Ying14ICC} (see Appendix \ref{app:ch}).

\subsubsection{Kronecker Channel Model}\label{sec:chk}
According to the simplified Kronecker model \cite{Ying14ICC, Gcong16TVT}, the channel vector $\pmb{g}_{ji}$ can be decomposed as
\begin{equation}\label{eq:kron_h}
\pmb{g}_{ji} = \zeta_{ji}\pmb{h}_{A,ji} \otimes \pmb{h}_{E,ji},
\end{equation}
where $\zeta_{ji}$ is defined as in (\ref{eq:pl}), while $\pmb{h}_{A, ji} \sim \mathcal{CN}(\pmb{0}_{N_{A,j} \times 1}, \pmb{R}_{h_{A,ji}})$ and $\pmb{h}_{E, ji} \sim \mathcal{CN}(\pmb{0}_{N_{E,j} \times 1}, \pmb{R}_{h_{E,ji}})$ are the $N_{A,j} \times 1$ azimuth and $N_{E,j} \times 1$ elevation components, respectively, with covariance matrices $\pmb{R}_{h_{A,ji}}\triangleq E[\pmb{h}_{A,ji}\pmb{h}^H_{A,ji}]$ and $\pmb{R}_{h_{E,ji}} \triangleq E[\pmb{h}_{E,ji}\pmb{h}^H_{E,ji}]$. These can be computed as $\left[\pmb{R}_{h_{A,ji}}\right]_{l,q} = \frac{1}{\sqrt{\gamma_5}}e^{-\frac{\gamma_3^2\cos^2\phi_{ji}}{2\gamma_5}}e^{j\frac{\gamma_2\cos\phi_{ji}}{\gamma_5}}$ $e^{-\frac{(\gamma_2\sigma_\phi\sin\phi_{ji})^2}{2\gamma_5}}$ and $\left[\pmb{R}_{h_{E,ji}}\right]_{k,p} = e^{j \frac{2\pi d_E}{\lambda}(p-k)\cos\theta_{ji}}e^{-\frac{1}{2}\left(\sigma_\theta\frac{2\pi d_E}{\lambda}\right)^2(p-k)^2\sin^2\theta_{ji}}$, where $[\cdot]_{l,q}$ denotes the $(l,q)$th entry of the argument matrix and the parameters $\{\gamma_k\}_{k \in \{1, \dots, 7\}}$ are defined  in Appendix \ref{app:ch} \cite{Ying14ICC}. The Kronecker model has been reported to be a valid approximation for propagation scenarios characterized by scatters distributed in the vicinity of BS and MS \cite{Gershman05}. We note that the correlated channel model includes the Kronecker channel model as a special case.

\subsection{Uplink Signal Model}\label{sec:sig_ul}
In this section, we model the pilot signals received at the BS $j \in \mathcal{N}_B$ during uplink training and the estimated channels of all the MSs at the BSs by accounting for the low-resolution ADCs of the BSs as in \cite{Mezghani12ISIT, Stein13ICASSP, Li16Arxiv}.  
 
In each uplink pilot slot of any band $j$, the MSs simultaneously transmit orthogonal pilot signals $\bar{\pmb{s}}^p_{ji} \in \mathbb{C}^{\bar{n}_p \times 1} = [\bar{s}^p_{ji}(1) \cdots \bar{s}^p_{ji}(\bar{n}_p)]^T$ of duration $\bar{n}_p  \ge N_M$ symbols with normalized energy $\|\bar{\pmb{s}}^{p}_{ji}\|^2/\bar{n}_p = 1$ for all $i \in \mathcal{N}_M$. 

To analyze the impact of quantization resulting from the low-resolution ADCs, as in \cite{Mezghani12ISIT, Stein13ICASSP, Li16Arxiv}, we leverage the Bussgang theorem \cite{Bussgang52}. Accordingly, we write the $N_j \times \bar{n}_p$ output of the ADCs at BS $j$ as  
\begin{equation}\label{eq:Bussgang_UL_re}
\bar{\pmb{U}}^p_{j} = \mathcal{Q}_{B_j}\left( \sum_{i \in \mathcal{N}_M}\sqrt{\bar{P}}\pmb{g}_{ji}\bar{\pmb{s}}^{p H}_{ji} + \bar{\pmb{Z}}^p_{j} \right) = (1-D_j)\sum_{i \in \mathcal{N}_M}\sqrt{\bar{P}}\pmb{g}_{ji}\bar{\pmb{s}}^{p H}_{ji} + \tilde{\pmb{Q}}^p_{j}, 
\end{equation}
where $\bar{P}$ is a transmit power per MS for uplink training; $\bar{\pmb{Z}}^p_{j} = [\bar{\pmb{z}}^{p}_{j}(1) \cdots \bar{\pmb{z}}^{p}_{j}(\bar{n}_p)]$ is the $N_j \times \bar{n}_p$ noise matrix, which consists of independent and identically distributed (i.i.d.) $\mathcal{CN}(0, N_0)$ entries; $D_j$ is tabulated in \cite{Max60IRE} for different values of the resolution $B_j$; and $\tilde{\pmb{Q}}^p_{j} = [\tilde{\pmb{q}}^{p}_{j}(1) \cdots \tilde{\pmb{q}}^{p}_{j}(\bar{n}_p)]$ is the $N_j \times \bar{n}_p$ noise matrix including the channel and quantization noise. Each $N_j \times 1$ noise vector $\tilde{\pmb{q}}^p_{j}(l)$ has a diagonal correlation matrix $\pmb{R}_{\tilde{q}^p_{j}} = (1-D_j)(N_0 \pmb{I}_{N_j \times N_j} + D_j\bar{P}\text{diag}\{\sum_{i \in \mathcal{N}_M}\pmb{R}_{g_{ji}}\})$, with $\text{diag}\{\cdot\}$ being a diagonal matrix whose diagonal elements are the same as the argument matrix, while off-diagonal elements are equal to zero. The effective noise matrix $\tilde{\pmb{Q}}^p_{j}$ is uncorrelated but not independent of the desired signal in (\ref{eq:Bussgang_UL_re}), and it is also not Gaussian. 

Using the linear minimum mean-square-error (LMMSE) approach, the channel $\pmb{g}_{ji}$ between BS $j$ and MS $i$ can be estimated  based on the received signal (\ref{eq:Bussgang_UL_re}), yielding the estimate \cite{Hassibi03TIT}
\begin{equation} \label{eq:hat_G}
\hat{\pmb{g}}_{ji} = \frac{1}{(1-D_j)\bar{n}_p\sqrt{\bar{P}}}\pmb{R}_{g_{ji}}\left( \pmb{R}_{g_{ji}} + \frac{1}{(1-D_j)^2\bar{n}_p\bar{P}}\pmb{R}_{\tilde{q}^p_j} \right)^{-1}\bar{\pmb{U}}^p_j\bar{\pmb{s}}^p_{ji}.
\end{equation}
From the orthogonality property of the LMMSE \cite{Hassibi03TIT}, the channel vector $\pmb{g}_{ji}$ can be written as a function of the estimate (\ref{eq:hat_G}) as $\pmb{g}_{ji} = \hat{\pmb{g}}_{ji} + \pmb{\Delta}_{g_{ji}}$, where $\hat{\pmb{g}}_{ji}$ and $\pmb{\Delta}_{g_{ji}}$ are zero-mean and uncorrelated with covariance matrices $\pmb{R}_{\hat{g}_{ji}}  
= \pmb{R}_{g_{ji}}\left( \pmb{R}_{g_{ji}} + \frac{1}{(1-D_j)^2\bar{n}_p\bar{P}}\pmb{R}_{\tilde{q}^p_j} \right)^{-1}\pmb{R}^H_{g_{ji}}$ and $\pmb{R}_{\Delta_{g_{ji}}}= \frac{1}{(1-D_j)^2\bar{n}_p\bar{P}}\pmb{R}_{g_{ji}}\left( \pmb{R}_{g_{ji}} + \frac{1}{(1-D_j)^2\bar{n}_p\bar{P}}\pmb{R}_{\tilde{q}^p_j} \right)^{-1}\pmb{R}^H_{\tilde{q}^p_j}$.  
\subsection{Downlink Signal Model}\label{sec:sig}
Here, we describe the signal model for downlink pilot and data transmission. In a manner similar to the treatment of uplink training, we take into account the impact of the limited resolution DACs available at the BSs.  

In both downlink training and data phases, due to the per-antenna low-resolution DAC at the BS $j$, the transmitted signals are quantized before transmission. The discrete-time transmitted signals are and given as \cite{Bussgang52}
\begin{equation}\label{eq:x_Q}
\pmb{u}^\nu_j(l) = \mathcal{Q}_{B_j}(\pmb{W}_{j}\pmb{s}^\nu_{j}(l)) =  (1-D_j)\sum_{i \in \mathcal{N}_M}\pmb{w}_{ji}s^\nu_{ji}(l) + \pmb{q}^\nu_{j}(l),
\end{equation}         
where the superscript $\nu \in \{p, d\}$ indicates the training, or pilot ($p$), phase and the data ($d$) phase; $\pmb{W}_j = [\pmb{w}_{j1} \cdots \pmb{w}_{jN_M}] \in \mathbb{C}^{N_j \times N_M}$ is the beamforming matrix, with $\pmb{w}_{ji}$ being the $N_j \times 1$ beamforming vector employed by BS $j$ to communicate with MS $i$ in the allocated band; and $\pmb{s}^\nu_j(l) = [s^\nu_{j1}(l) \cdots$ $s^\nu_{jN_M}(l)]^T \in \mathbb{C}^{N_M \times 1}$ with $s^p_{ji}(l)$, for $l=1, \dots, n_p$, and $s^d_{ji}(l)$, for $l=1, \dots, n_d$, are pilot symbols and data symbols, respectively, which are assumed to be i.i.d. $\mathcal{CN}(0,1)$ variables \cite{GamalBook}. In (\ref{eq:x_Q}), Bussgang theorem \cite{Bussgang52} is applied to model quantization, where $D_j$ is defined as in (\ref{eq:Bussgang_UL_re}), and $\pmb{q}^\nu_{j}(l)$ is the uncorrelated zero-mean quantization error vector with diagonal correlation matrix 
\begin{equation}\label{eq:Rqj}
\pmb{R}_{q_{j}}(\pmb{W}_j) = D_j(1-D_j)\text{diag}\{\pmb{W}_j\pmb{W}_j^H\}.
\end{equation}

In this work, the beamforming vectors $\{\pmb{w}_{ji}\}$ are jointly designed based on the estimated CSI vectors $\{\hat{\pmb{g}}_{ji}\}$ in (\ref{eq:hat_G}). In practice, this design can be carried out at a central unit (not shown in Fig. \ref{fig:sys}) that collects all the estimates from the BSs. We consider two beamforming design approaches for the vectors $\{\pmb{w}_{ji}\}$, namely: ($i$) \textit{unstructured design}, in which no constraints are imposed on the structure of the vectors $\{\pmb{w}_{ji}\}$; ($ii$) \textit{structured design} based on the Kronecker parameterization 
\begin{equation}\label{eq:w}
\pmb{w}_{ji} = \pmb{w}_{A, ji} \otimes \pmb{w}_{E, ji},
\end{equation}
with $\pmb{w}_{A, ji}$ being its $N_{A, j} \times 1$ azimuth component and $\pmb{w}_{E, ji}$ being its $N_{E, j} \times 1$ elevation component. This parameterization is particularly well studied for the scenarios in which the channel is well described by the Kronecker channel model and that it has been previously considered in \cite{Ying14ICC, Gcong16TVT}. It has the advantage that it decreases the number of parameters to be designed from $N_M \sum_{j \in \mathcal{N}_B}N_j =N_M \sum_{j \in \mathcal{N}_B}N_{A, j}N_{E, j}$ to $N_M \sum_{j \in \mathcal {N}_B}(N_{A, j}+ N_{E, j})$, hence significantly reducing the complexity of the design problem, specially in multiuser massive MIMO systems. We will study unstructured beamforming design under the general correlated channel model, and Kronecker beamforming design assuming a Kronecker channel model in order to obtain lower-complexity solutions. 

The BSs' antennas apply the analog filter $p(t)$ prior to transmission, where $p(t)$ is a unitary-energy Nyquist pulse whose Fourier transform is $P(f)$. As a result, the continuous-time signal $y^\nu_{ji}(t)$ received at MS $i$ during the pilot phase for $\nu=p$ or during the data phase for $\nu=d$ can be expressed as 
\begin{equation}\label{eq:y}
y^\nu_{ji}(t) = \pmb{g}^T_{ji}\pmb{u}^\nu_{j}(t-\tau_{ji}) + z^\nu_{ji}(t), 
\end{equation}   
where $\pmb{u}^\nu_{j}(t) = \sum_{l =1}^{n_\nu} \pmb{u}^\nu_{j}(l)p(t-(l-1)T_s)$ is the modulated transmitted signal, with $T_s$ being the symbol period; $\tau_{ji}=d_{ji}/c$ is the effective propagation delay between BS $j$ and MS $i$, with $c$ being the propagation speed; and the noise $z^\nu_{ji}(t)$ is complex white Gaussian with zero mean and two-sided power spectral density $N_0$. Each MS $i$ estimates its position $\pmb{p}_{M, i}$ as well as the downlink channels based on the signals $y^p_{ji}(t)$ received during downlink training phase, and then decodes the signals $y^d_{ji}(t)$ received during downlink data phase based on the available CSI obtained via downlink training. The consideration of the continuous-time signal (\ref{eq:y}) will be useful below when considering the performance of localization. 

Each MS $i$ passes the received signal $y^\nu_{ji}(t)$ through a filter matched to $p(t)$. Assuming time synchronization at the MS \cite{GamalBook}, we can write the discrete-time received signal at MS $i$ in terms of the effective complex gain between BS $j$ and MS $i$ expressed in terms of the BSs' channel estimate $\hat{\pmb{g}}_{ji}$, which is defined as
\vspace{-0.5cm}
\begin{equation}\label{eq:effch}
\alpha^{(k)}_{ji}(\pmb{w}_{jk}) = (1-D_j)\hat{\pmb{g}}^T_{ji}\pmb{w}_{jk}.
\end{equation}
\vspace{-0.3cm}
This is done as follows:
\begin{subequations}\label{eq:y_dis_re}
\begin{eqnarray}
y^\nu_{ji}(l) &=& \pmb{g}^T_{ji}\pmb{u}^\nu_{j}(l) + z^\nu_{ji}(l)\label{eq:Bussgang_DL_re}\\
&=& (1-D_j)\sum_{k \in \mathcal{N}_M} \hat{\pmb{g}}^T_{ji}\pmb{w}_{jk}s^\nu_{jk}(l) + (1-D_j)\sum_{k \in \mathcal{N}_M}\pmb{\Delta}_{g_{ji}}^T\pmb{w}_{jk}s^\nu_{jk}(l) + \pmb{g}^T_{ji}\pmb{q}^\nu_{j}(l) + z^\nu_{ji}(l)\label{eq:y_dis_re_ch}\\
&=& \sum_{k \in \mathcal{N}_M}\alpha^{(k)}_{ji}(\pmb{w}_{jk})s^\nu_{jk}(l) + \tilde{z}^\nu_{ji}(l),\label{eq:y_dis_re_ch_noise}
\end{eqnarray}
\end{subequations}
for $l = 1, \dots, n_\nu$, where $z^\nu_{ji}(l) \sim \mathcal{CN}(0, N_0)$ is i.i.d. additive Gaussian noise; and $\tilde{z}^\nu_{ji}(l)$ is a non-Gaussian effective noise term, which includes channel estimation noise, distortion noise and channel noise with power
\begin{equation}\label{eq:effect_noise}
\sigma_{\tilde{z}_{ji}}^2(\pmb{W}_j) = \sum_{k \in \mathcal{N}_M}(1-D_j)^2\text{tr}\{\pmb{R}_{\Delta_{g_{ji}}}\pmb{w}_{jk}\pmb{w}_{jk}^H\} + D_j(1-D_j)\text{tr}\{\pmb{R}_{g_{ji}}\text{diag}\{\pmb{W}_j\pmb{W}_j^H\}\} + N_0.
\end{equation} 
Note that (\ref{eq:y_dis_re}) is written as a function of the CSI available at the BSs, so as to obtain expressions for the performance metrics of interest (see next section) that can be optimized at a central unit connected all BSs.

\section{Performance Metrics}\label{sec:metric}
As discussed, we are interested in designing beamforming vectors based on the estimated CSI available at the BSs so as to guarantee performance requirements in terms of both data transmission rate and localization accuracy for the downlink. In this section, we discuss the calculation of the performance criteria of achievable transmission rates and of localization accuracy.  

\subsection{Transmission Rate}\label{sec:rate}
Each MS $i$ decodes on any band $j$ the signal received from BS $j$. In order to evaluate achievable rates that can be used at the BSs for beamforming design, we treat the additive noise and interference in (\ref{eq:y_dis_re}) as Gaussian and independent of the signal. As proved in \cite{Medard00TIT}, this yields a lower bound on the achievable rate based on signal (\ref{eq:y_dis_re}). We note that the resulting rate is, strictly speaking, only achievable if the MS can estimate correctly the effective channel gain $\{\alpha^{(k)}_{ji}(\pmb{w}_{jk})\}_{i, k \in \mathcal{N}_M}$ in (\ref{eq:y_dis_re}). An additional noise term could be added in order to account for the channel estimation errors at the MSs, but this is not done here so as to avoid introducing more notation.

Treating the effective noise term $\sum_{k \in \mathcal{N}_M, k \neq, i}\alpha^{(k)}_{ji}(\pmb{w}_{jk})s^d_{jk}(l) + \tilde{z}^d_{ji}(l)$ in (\ref{eq:y_dis_re}), which includes interference from the undesired signals as well as the noise $\tilde{z}^d_{ji}(l)$, as Gaussian and independent of the useful signal, we obtain the rate
\begin{equation}\label{eq:rate}
C_{ji}(\pmb{W}_j) \triangleq \frac{n_d}{N_B n}\log_2\left(1+\frac{\left|\alpha^{(i)}_{ji}(\pmb{w}_{ji})\right|^2}{\sum_{k \in \mathcal{N}_M, k \neq i}\left|\alpha^{(k)}_{ji}(\pmb{w}_{jk})\right|^2 + \sigma_{\tilde{z}_{ji}}^2(\pmb{W}_j)}\right)  \,\, \text{ (bps/Hz) },
\end{equation}
where $n = n_p + n_d$.  
\subsection{Localization Accuracy}\label{sec:local}
MS $i$ estimates its position $\pmb{p}_{M, i}$ based on the received pilot signals $\pmb{y}^p_{i}=[(\pmb{y}^{p}_{1i})^T \cdots, (\pmb{y}^{p}_{N_Bi})^T]^T$, where $\pmb{y}^p_{ji}$ is the vector representation of the time-series signal $y^p_{ji}(t)$ in (\ref{eq:y}). In order to evaluate the localization accuracy, we adopt the standard squared position error (SPE) criterion \cite{Kay93Book, JSA14TVT, JSA15TVT, JSA16IET}, which is defined for each MS $i$ as 
\begin{equation}\label{eq:spe}
\text{SPE}(\pmb{W}, \pmb{p}_{M, i}, \hat{\pmb{g}}_i)=\mathbb{E}\left[\left.\left\|\hat{\pmb{p}}_{M, i}-\pmb{p}_{M, i}\right\|^2 \right|\pmb{p}_{M, i}, \hat{\pmb{g}}_i \right],
\end{equation}   
with $\hat{\pmb{p}}_{M, i}$ being the position estimate at MS $i$ and $\hat{\pmb{g}}_i = \{\hat{\pmb{g}}_{ji}\}_{j \in \mathcal{N}_B}$ being the channel estimates available at the BSs. The expectation in (\ref{eq:spe}) is conditioned on the unknown parameters including user's position and the channel estimates. We evaluate a lower bound on (\ref{eq:spe}) by considering the modified Cram\'{e}r-Rao bound (MCRB) \cite{Mengali94, Gini98} that is obtained as the trace of the inverse of the average equivalent Fisher information matrix (EFIM), when the average is over the pilot sequence. We specifically use the lower bound of the EFIM derived in \cite{Stein14SPL, Stein17TSP} by treating the additive noise in (\ref{eq:y_dis_re}) as Gaussian. Accordingly, the resulting performance metric for the SPE of MS $i$ as the MCRB is given as    
\begin{equation}\label{eq:crb}
\text{SPE}(\pmb{W}, \pmb{p}_{M, i}, \hat{\pmb{g}}_i) \le \rho(\pmb{W}, \pmb{p}_{M, i}, \hat{\pmb{g}}_i) \triangleq \text{tr}\left\{\pmb{J}^{-1}(\pmb{W}, \pmb{p}_{M, i}, \hat{\pmb{g}}_i)\right\},
\end{equation} 
where 
\vspace{-0.5cm}
\begin{eqnarray}\label{eq:efim}
\pmb{J}(\pmb{W}, \pmb{p}_{M, i}, \hat{\pmb{g}}_i) &=& \frac{8 \pi^2 n_p \beta^2}{c^2} \sum_{j \in \mathcal{N}_B} \frac{ \sum_{k \in \mathcal{N}_M}\left| \alpha^{(k)}_{ji}(\pmb{w}_{jk})\right|^2}{\sigma_{\tilde{z}_{ji}}^2(\pmb{W}_j) } \pmb{j}_{ji}\pmb{j}_{ji} ^T \nonumber\\
&=& \frac{8 \pi^2 n_p \beta^2}{c^2} \sum_{j \in \mathcal{N}_B} \frac{ \sum_{k \in \mathcal{N}_M}\left| \alpha^{(k)}_{ji}(\pmb{w}_{jk})\right|^2}{\sigma_{\tilde{z}_{ji}}^2(\pmb{W}_j) } \pmb{J}_{\phi, \theta} (\phi_{ji}, \theta_{ji}),
\end{eqnarray}
with $\beta=\{\int |fP(f)|^2 df\}^{1/2}$ being the effective bandwidth and $P(f)$ being the Fourier transform of the filter $p(t)$, $\pmb{j}_{ji}= \frac{1}{d_{ji}}(\pmb{p}_{M,i}-\pmb{p}_{B,j}) 
= [\sin\phi_{ji}\sin\theta_{ji} \,\,\,\, \cos\phi_{ji}\sin\theta_{ji} \,\,\,\, \cos\theta_{ji}]^T$ and $\pmb{J}_{\phi, \theta} (\phi, \theta) = \pmb{j}_{ji}\pmb{j}_{ji} ^T$. The detailed derivation of (\ref{eq:crb}) is provided in Appendix \ref{app:crb}.

\subsection{Total Transmit Power}\label{sec:pwr}
We conclude this section by evaluating the transmitted power for the BSs as a function of the beamforming vectors. Applying the Bussgang theorem \cite{Bussgang52} to the quantized precoded signals in (\ref{eq:x_Q}), the total transmit powers during training phase and data phase are given as  
\begin{equation}\label{eq:pwr}
\sum_{j \in \mathcal{N}_B}P_j(\pmb{W}_j) = \sum_{j \in \mathcal{N}_B}\mathbb{E}\left[\left\|\pmb{u}^\nu_{j}(l)\right\|^2\right]
= \sum_{j \in \mathcal{N}_B} \text{tr}\left\{(1-D_j)\pmb{W}_j \pmb{W}^H_j \right\}.
\end{equation}
Note that the transmit powers at BS $j$ during training phase and data phase are equal. 

\section{Unstructured Beamforming Design}\label{sec:beam_ideal}
In this section, we aim at optimizing the beamforming strategy by assuming the correlated channel model described in Section \ref{sec:chg} and without imposing any structure on the beamforming matrix. We will focus on two dual problems: ($i$) sum-rate maximization under localization accuracy and total transmit power constraints; and ($ii$) sum-SPE minimization under the data rate and total transmit power constraints. We here consider the average localization performance with respect to the MSs' positions, given that the MSs are spatially and randomly distributed within the given cube.   

\subsection{Problem Formulations}\label{sec:prob}
We denote the data rate and localization accuracy requirements for each MS $i$ as $C^{\min}_i$ and $\rho^{\max}_i$, respectively, while the constraint of total transmit power expenditure in the downlink is denoted as $P$. By using the performance metrics discussed in Section \ref{sec:metric}, namely the rate function $C_{ji}(\pmb{W}_j)$ in (\ref{eq:rate}), the SPE function $\rho(\pmb{W}, \pmb{p}_{M, i}, \hat{\pmb{g}}_i)$ in (\ref{eq:crb}) and the transmit power function $P_j(\pmb{W}_j)$ in (\ref{eq:pwr}), two design problems under study are formulated as   
\vspace{-0.2cm}
\begin{subequations}\label{eq:opt_rate}
\begin{eqnarray}
&& \hspace{-3.4cm} \text{(\bf{P1}: Sum-rate Maximization)}\hspace{0.5cm} \underset{\pmb{W}}{\text{maximize}} \hspace{0.5cm}\sum_{j \in \mathcal{N}_B, i \in \mathcal{N}_M} C_{ji}(\pmb{W}_j) \label{eq:P1_rate}\\
&& \hspace{3cm}  \text{s.t.} \hspace{0.5cm} \mathbb{E}_{\pmb{p}_{M, i}}\left[\rho(\pmb{W}, \pmb{p}_{M, i}, \hat{\pmb{g}}_i)\right] \le \rho^{\max}_i, \hspace{0.2cm} \text{for}\,\, \text{all}\,\, i \in \mathcal{N}_M \label{eq:P1_local}\\
&& \hspace{3.9cm} \sum_{j \in \mathcal{N}_B} P_j(\pmb{W}_j)  \le P,
\end{eqnarray}
\end{subequations}  
\vspace{-0.2cm}
and
\begin{subequations}\label{eq:opt_local}
\begin{eqnarray}
&& \hspace{-4.9cm} \text{(\bf{P2}: Sum-SPE Minimization)}\hspace{0.5cm} \underset{\pmb{W}}{\text{minimize}} \hspace{0.5cm} \sum_{i \in \mathcal{N}_M}\mathbb{E}_{\pmb{p}_{M, i}}\left[\rho(\pmb{W}, \pmb{p}_{M, i}, \hat{\pmb{g}}_i)\right] \label{eq:P2_local} \\
&& \hspace{1.5cm}  \text{s.t.} \hspace{0.5cm} \sum_{j \in \mathcal{N}_B} C_{ji}(\pmb{W}_j) \ge C^{\min}_i, \hspace{0.2cm} \text{for}\,\, \text{all}\,\, i \in \mathcal{N}_M \label{eq:P2_rate}\\
&& \hspace{2.5cm} \sum_{j \in \mathcal{N}_B} P_j(\pmb{W}_j) \le P.
\end{eqnarray}
\end{subequations}  
The problems {\bf{P1}} and {\bf{P2}} are challenging to solve due to the expectation in the localization error performance (\ref{eq:P1_local}) and (\ref{eq:P2_local}) with respect to the unknown users' locations, and to the non-convexity of both rate and localization criteria (\ref{eq:rate}) and (\ref{eq:crb}). To address the first issue, we approximate the expectation in (\ref{eq:P1_local}) and (\ref{eq:P2_local}) by using the sample average approximation method \cite{Shapiro09SIAM}. Accordingly, we estimate the expectation of $\mathbb{E}_{\pmb{p}_{M, i}}[\rho(\pmb{W}, \pmb{p}_{M, i}, \hat{\pmb{g}}_i)]$ with respect to the user position as 
\begin{equation}\label{eq:SAA}
\mathbb{E}_{\pmb{p}_{M, i}}[\rho(\pmb{W}, \pmb{p}_{M, i}, \hat{\pmb{g}}_i)] \approx \frac{1}{N_s}\sum_{m = 1}^{N_s}\rho(\pmb{W}, \pmb{p}_{M, i, m}, \hat{\pmb{g}}_i),
\end{equation}   
for $i \in \mathcal{N}_M$, where $\pmb{p}_{M, i, 1}, \pmb{p}_{M, i, 2}, \dots, \pmb{p}_{M, i, N_s}$ are $N_s$ independent realizations of the MS $i$'s position. The latter is assumed to be uniformly distributed in the $\Delta_x \times \Delta_y \times \Delta_z$ cube as described in Section \ref{sec:sys} \footnote{In principle, based on the channel estimates $\{\hat{\pmb{g}}_i\}$, the optimizer could restrict the uncertainty area for the user to a smaller volume. The assumed uniform distribution can hence be thought of as providing a worst-case performance.}. To tackle the second issue, we apply rank relaxation and the successive convex approximation (SCA) method introduced in \cite{Scutari14Arxiv, Scutari16ArxivII} as detailed in the following.    
      
\subsection{Sum-Rate Maximization}\label{sec:prob_R}
In this section, we elaborate on the solution of sum-rate maximization problem {\bf{P1}} with the stochastic approximation (\ref{eq:SAA}). To this end, we start by reformulating the problem with respect to the beamforming covariance matrices $\pmb{\Omega} = \{\pmb{\Omega}_{ji}\}_{j \in \mathcal{N}_B, i \in \mathcal{N}_M}$ with $\pmb{\Omega}_{ji}=\pmb{w}_{ji}\pmb{w}_{ji}^H$. Accordingly, the elements of the diagonal covariance matrix $\pmb{R}_{q_{j}}(\pmb{W}_j)$ of the distortion noise in (\ref{eq:Rqj}) during both training phase and data phase can be expressed in terms of $\pmb{\Omega}$ as
\begin{equation}
\left[\pmb{R}_{q_{j}}(\pmb{W}_j)\right]_{n,n} = D_j(1-D_j)\sum_{i \in \mathcal{N}_M} \text{tr}\{\pmb{E}_n \pmb{\Omega}_{ji} \pmb{E}_n^T\}, 
\end{equation}
for $n = 1,\dots,N_j$, where we have defined the matrix $\pmb{E}_n = \pmb{e}_n\pmb{e}_n^T$, with $\pmb{e}_n$ being a vector whose $n$th entry equals to $1$ and the rest equal to zero. Similarly, the effective power gain $|\alpha^{(k)}_{ji,k}(\pmb{w}_{jk})|^2$ in (\ref{eq:effch}) can be written as $\xi^{(k)}_{ji}(\pmb{\Omega}_{jk}) = (1-D_j)^2\hat{\pmb{g}}^H_{ji}\pmb{\Omega}_{jk}\hat{\pmb{g}}_{ji}$. Introducing the auxiliary variables $\pmb{\kappa} = \{\kappa_{jn}\}_{j \in \mathcal{N}_B, n = 1, \dots, N_j}$ for each BS antenna, where $\kappa_{jn} \in \mathbb{R}$ and variables $\pmb{\chi} = \{\chi_{ji}\}_{j \in \mathcal{N}_B, i \in \mathcal{N}_M}$ for each BS-MS pair, where $\chi_{ji} \in \mathbb{R}$, a rank-relaxed version of problem {\bf{P1}} based on the empirical approximation (\ref{eq:SAA}) can be written as  
\begin{subequations}\label{eq:opt_rate1} 
\begin{eqnarray}
&& \hspace{-1.7cm} \text{(\bf{P1-1}):}\hspace{0.2cm} \underset{\pmb{\Omega}, \pmb{\kappa}, \pmb{\chi}}{\text{maximize}} \hspace{0.2cm} \frac{n_d}{N_Bn} \sum_{j \in \mathcal{N}_B, i \in \mathcal{N}_M} \log_2\left( \sum_{k \in \mathcal{N}_M} \xi^{(k)}_{ji}(\pmb{\Omega}_{jk}) + N_0 + \sum_{k \in \mathcal{N}_M}(1-D_j)^2\text{tr}\{\pmb{R}_{\Delta_{g_{ji}}}\pmb{\Omega}_{jk}\}\right.\nonumber\\
&& \hspace{5cm}\left. +  \sum_{n=1}^{N_j} \kappa_{jn}\text{tr}\left\{\pmb{E}_n\pmb{R}_{g_{ji}}\pmb{E}_n^T\right\} \right) - \log_2\left( \sum_{k \in \mathcal{N}_M, k \neq, i} \xi^{(k)}_{ji}(\pmb{\Omega}_{jk}) + N_0  \right.\nonumber\\
&& \hspace{5cm}  \left. + \sum_{k \in \mathcal{N}_M}(1-D_j)^2\text{tr}\{\pmb{R}_{\Delta_{g_{ji}}}\pmb{\Omega}_{jk}\} +  \sum_{n=1}^{N_j} \kappa_{jn}\text{tr}\left\{\pmb{E}_n\pmb{R}_{g_{ji}}\pmb{E}_n^T\right\} \right) \label{eq:R_obj1}\\
&& \hspace{-1cm}  \text{s.t.} \hspace{0.2cm} \frac{1}{N_s}\sum_{m = 1}^{N_s}\text{tr}\left\{\left(\frac{8 \pi^2 n_p \beta^2}{c^2} \sum_{j=1}^{N_B} \pmb{J}_{\phi, \theta} (\phi_{ji,m}, \theta_{ji,m}) \chi_{ji} \right)^{-1}\right\} \le \rho^{\max}_i, \hspace{0.2cm} i \in \mathcal{N}_M \label{eq:R_CRB1}\\
&& \hspace{-0.7cm} \sum_{j \in \mathcal{N}_B, i \in \mathcal{N}_M} (1-D_j)\text{tr}\left\{\pmb{\Omega}_{ji}\right\} \le P \label{eq:R_pt1}\\
&& \hspace{-0.7cm} \frac{ \sum_{k \in \mathcal{N}_M} \xi^{(k)}_{ji}(\pmb{\Omega}_{jk})}{N_0 + \sum_{k \in \mathcal{N}_M}(1-D_j)^2\text{tr}\{\pmb{R}_{\Delta_{g_{ji}}}\pmb{\Omega}_{jk}\}  +  \sum_{n=1}^{N_j} \kappa_{jn}\text{tr}\left\{\pmb{E}_n\pmb{R}_{g_{ji}}\pmb{E}_n^T\right\}} \ge \chi_{ji}, \hspace{0.2cm} j \in \mathcal{N}_B, i \in \mathcal{N}_M \label{eq:R_chi1}\\
&& \hspace{-0.7cm} D_j(1-D_j)\sum_{i \in \mathcal{N}_M}\text{tr}\left\{\pmb{E}_n\pmb{\Omega}_{ji}\pmb{E}_n^T\right\} \le \kappa_{jn}, \hspace{0.2cm} j \in \mathcal{N}_B \hspace{0.2cm} \text{and} \hspace{0.2cm} n=1, \dots, N_j \label{eq:R_qq1}\\
&& \hspace{-0.7cm} \pmb{\Omega}_{ji} \succeq 0, \hspace{0.2cm} j \in \mathcal{N}_B, i \in \mathcal{N}_M \label{eq:R_spd1}\\
&& \hspace{-0.7cm} \chi_{ji} \ge 0, \hspace{0.2cm} j \in \mathcal{N}_B, i \in \mathcal{N}_M, \label{eq:R_chi_plus1}
\end{eqnarray}
\end{subequations}
where $\phi_{ji,m}$ and $\theta_{ji,m}$ are azimuth and elevation angle between the $N_s$ independent realizations of the MS $i$'s position and BS $j$. The rank-relaxation in (\ref{eq:opt_rate1}) is obtained by dropping the constraint $\text{rank}(\pmb{\Omega}_{ji}) = 1$ for $j \in \mathcal{N}_B$ and $i \in \mathcal{N}_M$. Note that the equivalence between problem {\bf{P1}} and {\bf{P1-1}} under the rank relaxation follows from the fact that, at an optimum, inequalities (\ref{eq:R_chi1}) and (\ref{eq:R_qq1}) can be assumed to hold with equality without loss of optimality. The rank-relaxed problem {\bf{P1-1}} is still not convex owing to the presence of the non-convex objective function (\ref{eq:R_obj1}) and non-convex constraints (\ref{eq:R_chi1}). To resolve this problem, we apply the SCA method introduced in \cite{Scutari14Arxiv, Scutari16ArxivII}, which yields an iterative algorithm that is guaranteed to converge to a stationary point of the original non-convex problem under suitable conditions. This algorithm solves a sequence of strongly convex problems obtained as a local approximation of the original non-convex problem. In order to develop the SCA-based algorithm, we use the following lemmas. 

\textit{Lemma 1} ([43, Example 7]): Consider a non-concave utility function $U(\pmb{x}) = \sum_{i = 1}^{I}f_i(\pmb{x})$ with $f_i(\pmb{x}) = f^{+}_i(\pmb{x}) - f^{-}_i(\pmb{x})$ being a difference of concave (DC) function, where $f^{+}_ i(\pmb{x})$ and $f^{-}_i(\pmb{x})$ are  concave and continuously differentiable. Then, for any $\pmb{y}$ in the domain of $U(\pmb{x})$, a concave approximant of $U(\pmb{x})$ that has the properties required by the SCA algorithm \cite[Assumption 2]{Scutari14Arxiv} is given as
\begin{equation}\label{eq:SCA_obj}
\hat{U}(\pmb{x}; \pmb{y}) \triangleq \sum_{i = 1}^{I} f^{+}_i(\pmb{x}) - \nabla_{\pmb{x}}f^{-}_i(\pmb{y})^T(\pmb{x} - \pmb{y}) - \frac{\tau_i}{2}(\pmb{x}_i - \pmb{y}_i)^T\pmb{H}_i(\pmb{y})(\pmb{x}_i - \pmb{y}_i) ,
\end{equation}
where $\tau_i > 0$ is a positive constant ensuring that (\ref{eq:SCA_obj}) is strongly concave and $\pmb{H}_i(\pmb{y})$ is any uniformly positive definite matrix. 

\textit{Lemma 2} ([43, Example 4]): Consider a non-convex constraint $g(\pmb{x}_1, \pmb{x}_2) \le 0$, where $g(\pmb{x}_1, \pmb{x}_2) = h_1(\pmb{x}_1)h_2(\pmb{x}_2)$ and $h_1(\pmb{x}_1)$ and $h_2(\pmb{x}_2)$ are convex and non-negative. Then, for any $(\pmb{y}_1, \pmb{y}_2)$ in the domain of $g(\pmb{x}_1, \pmb{x}_2)$, a convex approximant that satisfies the conditions \cite[Assumption 3]{Scutari14Arxiv} required by the SCA algorithm is given as
\begin{eqnarray}\label{eq:SCA_con1}
\bar{g}(\pmb{x}_1, \pmb{x}_2; \pmb{y}_1, \pmb{y}_2) &\triangleq& \frac{1}{2}\left(h_1(\pmb{x}_1) + h_2(\pmb{x}_2)\right)^2 - \frac{1}{2}\left(h_1^2(\pmb{y}_1) + h_2^2(\pmb{y}_2)\right) \nonumber\\
&& \hspace{1.5cm} - h_1(\pmb{y}_1)\nabla_{\pmb{x}_1} h_1^T(\pmb{y}_1)(\pmb{x}_1 - \pmb{y}_1) - h_2(\pmb{y}_2)\nabla_{\pmb{x}_2}h_2^{T}(\pmb{y}_2)(\pmb{x}_2 - \pmb{y}_2).  
\end{eqnarray} 

Using Lemma 1 for the objective function (\ref{eq:R_obj1}), Appendix \ref{app:P1} shows that a strongly concave surrogate function can be obtained as
\begin{eqnarray}\label{eq:hatC} 
&&\hspace{-2cm}\hat{C}(\pmb{v}; \pmb{v}(t)) \triangleq  \frac{n_d}{N_Bn} \left(\sum_{j, i} f^{+}_{ji}(\pmb{v}) - \sum_{j , i}\nabla_{\pmb{v}} f^{-}_{ji}(\pmb{v}(t))^T(\pmb{v} - \pmb{v}(t))\right) \nonumber\\
&& \hspace{2cm} - \sum_{j, i} \frac{\tau_{\Omega_{ji}}}{2}\left\|\pmb{\Omega}_{ji} - \pmb{\Omega}_{ji}(t)\right\|_F^2  -  \sum_{j, n} \frac{\tau_{\kappa_{jn}}}{2}(\kappa_{jn}-\kappa_{jn}(t))^2,
\end{eqnarray}
where $f^{+}_{ji}(\pmb{v})$ and $f^{-}_{ji}(\pmb{v})$ are defined as (\ref{eq:R_DC_func});
\begin{eqnarray}\label{eq:fm_der}
&&\hspace{-1.3cm}\nabla_{\pmb{v}} f^{-}_{ji}(\pmb{v}(t))^T(\pmb{v} - \pmb{v}(t)) = \frac{1}{\ln 2} \left( \sum_{k \in \mathcal{N}_M, k \neq, i} (1-D_j)^2\hat{\pmb{g}}^H_{ji}(\pmb{\Omega}_{jk} - \pmb{\Omega}_{jk}(t))\hat{\pmb{g}}_{ji} \right.\nonumber\\
&&\hspace{-0.8cm} \left. + \sum_{k \in \mathcal{N}_M}(1-D_j)^2\text{tr}\{\pmb{R}_{\Delta_{g_{ji}}}(\pmb{\Omega}_{jk} - \pmb{\Omega}_{jk}(t))\} +  \sum_{n=1}^{N_j} (\kappa_{jn} - \kappa_{jn}(t)) \text{tr}\left\{\pmb{E}_n\pmb{R}_{g_{ji}}\pmb{E}_n^T\right\}\right) \nonumber\\
&& \hspace{-0.8cm} \div \left( \sum_{k \in \mathcal{N}_M, k \neq, i} \xi^{(k)}_{ji}(\pmb{\Omega}_{jk}(t)) +  N_0 + \sum_{k \in \mathcal{N}_M}(1-D_j)^2\text{tr}\{\pmb{R}_{\Delta_{g_{ji}}}\pmb{\Omega}_{jk}(t)\} +  \sum_{n=1}^{N_j} \kappa_{jn}(t)\text{tr}\left\{\pmb{E}_n\pmb{R}_{g_{ji}}\pmb{E}_n^T\right\} \right); 
\end{eqnarray}
and we have defined the set of optimization variables $\pmb{v} = (\pmb{\Omega}, \pmb{\kappa}, \pmb{\chi})$ and $\pmb{v}(t) = (\pmb{\Omega}(t), \pmb{\kappa}(t), \pmb{\chi}(t))$ as the $t$th iterate of the SCA algorithm. Furthermore, applying Lemma 2 to constraint (\ref{eq:R_chi1}), we obtain the following strongly concave approximation of the problem {\bf{P1-1}} for a given a feasible solution $\pmb{v}(t)$ as 
\begin{subequations}\label{eq:opt_rate2} 
\begin{eqnarray}
&& \hspace{-1.8cm} \text{(\bf{P1-2}):}\hspace{0.2cm} \underset{\pmb{v}}{\text{maximize}} \hspace{0.6cm} \hat{C}(\pmb{v}; \pmb{v}(t)) \label{eq:R_obj2}\\
&& \hspace{0.3cm}  \text{s.t.} \hspace{0.3cm} \bar{g}_{ji}(v_1, \pmb{v}_2; v_1(t), \pmb{v}_2(t)) \le \sum_{k \in \mathcal{N}_M} \xi^{(k)}_{ji}(\pmb{\Omega}_{jk}) - \chi_{ji}N_0, \hspace{0.2cm} j \in \mathcal{N}_B, i \in \mathcal{N}_M \label{eq:R_chi2}\\
&& \hspace{1.1cm}\text{(\ref{eq:R_CRB1}), (\ref{eq:R_pt1}), (\ref{eq:R_qq1})-(\ref{eq:R_chi_plus1})},
\end{eqnarray}
\end{subequations} 
where 
\begin{eqnarray}\label{eq:EFIM_chi_upper}
&&\hspace{-1.2cm}\bar{g}_{ji}(v_1, \pmb{v}_2; v_1(t), \pmb{v}_2(t)) = \nonumber\\
&&\hspace{-1.2cm} \frac{1}{2}\left(h_{1, ji}(v_1) + h_{2, ji}(\pmb{v}_2)\right)^2 - \frac{1}{2}\left(h_{1, ji}^2(v_1(t)) + h_{2, ji}^2(\pmb{v}_2(t))\right) - h_{1,ji}(v_1(t))(\chi_{ji} - \chi_{ji}(t) )   \nonumber\\
&&\hspace{-1.2cm} - h_{2, ji}(\pmb{v}_2(t))\left( \sum_{k \in \mathcal{N}_M}(1-D_j)^2\text{tr}\{\pmb{R}_{\Delta_{g_{ji}}} (\pmb{\Omega}_{jk} - \pmb{\Omega}_{jk}(t))\}  +  \sum_{n=1}^{N_j} (\kappa_{jn} - \kappa_{jn}(t))\text{tr}\left\{\pmb{E}_n\pmb{R}_{g_{ji}}\pmb{E}_n^T\right\}\right);  
\end{eqnarray}  
$h_{1, ji}(v_1) = \chi_{ji}$; $h_{2, ji}(\pmb{v}_2) = \sum_{k \in \mathcal{N}_M}(1-D_j)^2\text{tr}\{\pmb{R}_{\Delta_{g_{ji}}}\pmb{\Omega}_{jk}\}  +  \sum_{n=1}^{N_j} \kappa_{jn}\text{tr}\left\{\pmb{E}_n\pmb{R}_{g_{ji}}\pmb{E}_n^T\right\}$; $v_1 = \chi_{ji}$; $\pmb{v}_2 = \{\pmb{\Omega}_{j1}, \dots, \pmb{\Omega}_{jN_M}, \kappa_{j1}, \dots, \kappa_{jN_j}\}$; $v_1(t) = \chi_{ji}(t)$ and $\pmb{v}_2 = \{\pmb{\Omega}_{j1}(t), \dots, \pmb{\Omega}_{jN_M}(t), \kappa_{j1}(t), \dots, \kappa_{jN_j}(t)\}$. Additional details for the derivation of (\ref{eq:opt_rate2}) can be found in Appendix \ref{app:P1}. The problem {\bf{P1-2}} has a unique solution denoted by $\hat{\pmb{v}}(\pmb{v}(t))$. Using problem {\bf{P1-2}}, the proposed SCA-based algorithm is summarized in Algorithm \ref{al1}. Algorithm \ref{al1} obtains a solution $(\pmb{\Omega}^\opt, \pmb{\kappa}^\opt, \pmb{\chi}^\opt)$ using SCA, and then computes a feasible beamforming vectors $\{\pmb{w}^\opt_{ji}\}$ from the covariance matrices $\{\pmb{\Omega}^\opt_{ji}\}$ using standard rank-reduction method coupled with the scaling method \cite{Luo10SPM}. It is noted that the rate and the EFIM functions are monotonic with respect to the scaling factor $\epsilon_s$, which entails that Algorithm \ref{al1} provides a feasible solution for the original problem {\bf{P1}}.   
  
\linespread{1}
\begin{algorithm} [t] 
\begin{algorithmic}
\caption{Unstructured beamforming design for sum-rate maximization \& sum-SPE minimization} \label{al1}
\State {\textbf{Initialization}}: Initialize $\pmb{v}(0) = \{\pmb{\Omega}(0), \pmb{\kappa}(0), \pmb{\chi}(0)\}$ and fix parameters $\tau_{\Omega_{ji}}, \tau_{\kappa_{jn}} > 0$ for $j \in \mathcal{N}_B$, $i \in \mathcal{N}_M$, $n = \{1, \dots, N_j\}$ and $\epsilon_v, \epsilon_s > 0$. Set $t=0$.
\State  {\textbf{Repeat}} 
\State \indent 1. Compute $\hat{\pmb{v}}(\pmb{v}(t))$ using {\bf{P1-2}} for sum-rate maximization (or {\bf{P2-2}} for sum-SPE minimization);
\State \indent 2. Set $\pmb{v}(t+1) = \pmb{v}(t) + \epsilon(t)(\hat{\pmb{v}}(\pmb{v}(t)) - \pmb{v}(t))$ for some $\epsilon(t) \in (0, 1]$;
\State \indent 3. $t \gets t+1$;
\State \indent 4. If $\left\|\pmb{v}(t) - \pmb{v}(t-1)\right\|< \epsilon_v $, stop. Otherwise, go to step 1.
\State {\textbf{end}}
\State $(\pmb{\Omega}^\opt, \pmb{\kappa}^\opt, \pmb{\chi}^\opt) \gets \hat{\pmb{v}}(\pmb{v}(t))$.
\State {\bf{Rank reduction}}: Extract $\hat{\pmb{w}}_{ji}=\sqrt{\lambda_{\max}(\pmb{\Omega}^\opt_{ji})}\pmb{v}_{\max}(\pmb{\Omega}^\opt_{ji})$ for all $j \in \mathcal{N}_B$ and $i \in \mathcal{N}_M$, where  $\lambda_{\max}(\pmb{\Omega}^\opt_{ji})$ and $\pmb{v}_{\max}(\pmb{\Omega}^\opt_{ji})$ are the maximum eigenvalue and the corresponding eigenvector of the beamforming matrix $\pmb{\Omega}^\opt_{ji}$, respectively.
\State {\bf{Scaling}}: Check whether $\hat{\pmb{w}}_{ji}$ is feasible for the sum-rate maximization (or the sum-SPE minimization) or not. If so, $\pmb{w}^\opt_{ji}=\hat{\pmb{w}}_{ji}$. Otherwise, rescale $\hat{\pmb{w}}_{ji} \leftarrow (1+\epsilon_s)\hat{\pmb{w}}_{ji}$ for any positive integer $\epsilon_s$ until $\hat{\pmb{w}}_{ji}$ is feasible. 
\end{algorithmic}
\end{algorithm}
\linespread{2}

\subsection{Sum-SPE Minimization}\label{sec:prob_CRB}
The sum-SPE minimization problem {\bf{P2}} with the stochastic approximation (\ref{eq:SAA}) in lieu of (\ref{eq:P2_local}) can be addressed by using rank relation and SCA in a manner similar to that detailed above for problem {\bf{P1}}. Specifically, by introducing the beamforming covariance matrices $\pmb{\Omega}$ and the auxiliary variables $\pmb{\kappa}$ and $\pmb{\chi}$ as in Section \ref{sec:prob_R} and following similar steps, we derive in Appendix \ref{app:P2} the SCA algorithm detailed in Algorithm \ref{al1}, where the strongly convex problem {\bf{P2-2}} is defined as    
\begin{subequations}\label{eq:opt_local2} 
\begin{eqnarray}
&& \hspace{-2.7cm} \text{(\bf{P2-2}):}\hspace{0.2cm} \underset{\pmb{v}}{\text{minimize}} \hspace{0.2cm} \sum_{i \in \mathcal{N}_M}\frac{1}{N_s}\sum_{m=1}^{N_s}\text{tr}\left\{\left(\frac{8 \pi^2 n_p \beta^2}{c^2} \sum_{j=1}^{N_B}\pmb{J}_{\phi, \theta} (\phi_{ji, m}, \theta_{ji, m}) \chi_{ji} \right)^{-1}\right\} \label{eq:L_obj2}\\
&& \hspace{-0.7cm} \text{s.t.} \hspace{0.7cm} \sum_{j \in \mathcal{N}_B}\bar{C}_{ji}(\pmb{v}; \pmb{v}(t)) \ge C^{\min}_i, \hspace{0.2cm} i \in \mathcal{N}_M \label{eq:L_rate2}\\
&& \hspace{0.5cm} \text{(\ref{eq:R_chi2}), (\ref{eq:R_pt1}), (\ref{eq:R_qq1})-(\ref{eq:R_chi_plus1})}.
\end{eqnarray}
\end{subequations} 
Problem {\bf{P2-2}} has a unique solution denoted by $\hat{\pmb{v}}(\pmb{v}(t))$. In {\bf{P2-2}}, we have defined the set of optimization variables for the rank-relaxed problem {\bf{P2-1}} in (\ref{eq:opt_local1}) for sum-SPE minimization {\bf{P2}} as $\pmb{v} = (\pmb{\Omega}, \pmb{\kappa}, \pmb{\chi})$ and $\pmb{v}(t) = (\pmb{\Omega}(t), \pmb{\kappa}(t), \pmb{\chi}(t))$ for the $t$th iterate within the feasible set of problem {\bf{P2-1}} and a concave lower bound $\bar{C}_{ji}(\pmb{v}; \pmb{v}(t))$ is derived as (\ref{eq:barC}) in Appendix \ref{app:P2} for the MS $i$'s achievable rate constraint (\ref{eq:P2_rate}). As for the sum-rate maximization, we extract the feasible beamforming vectors $\{\pmb{w}_{ji}^\opt\}$ from $\{\pmb{\Omega}^\opt_{ji}\}$ resulting from the solution $\hat{\pmb{v}}(\pmb{v}(t))$ with the standard rank-reduction method coupled with scaling method \cite{Luo10SPM}. Its convergence is established by the property of SCA method and monotonicity of the rate and the EFIM functions with respect to the scaling factor $\epsilon_s$ as discussed in Section \ref{sec:prob_R}.  

\section{Kronecker Beamforming Design}\label{sec:beam_kron}
In this section, we investigate a reduced-complexity beamforming design based on the parameterization of the Kronecker channel model (\ref{eq:kron_h}). We tackle both the sum-rate maximization ({\bf{P1}}) and the sum-SPE minimization ({\bf{P2}}). To this end, we apply a similar approach to that developed for unstructured beamforming with the caveat that we tackle alternately the optimization of the azimuth and elevation components by leveraging the decomposition of the FD channel model. The solution will also require to estimate the azimuth and elevation components of the channel, which will be done by using the solution called ``Nearest Kronecker product'' \cite{Golub12Book, Loan92Book}.

Based on the decomposition of beamforming vectors $\pmb{w}_{ji}$ in (\ref{eq:w}), we start by writing the beamforming covariance matrices $\pmb{\Omega}_{ji}$ in the form $\pmb{\Omega}_{ji} = \pmb{\Omega}_{A,ji} \otimes \pmb{\Omega}_{E,ji}$ with the rank-1 azimuth covariance matrices $\pmb{\Omega}_{A,ji} = \pmb{w}_{A,ji}\pmb{w}_{A,ji}^H$ and the elevation covariance matrices $\pmb{\Omega}_{E,ji} = \pmb{w}_{E,ji}\pmb{w}_{E,ji}^H$ for all $j \in \mathcal{N}_B$ and $i \in \mathcal{N}_M$. As mentioned, the azimuth components $\hat{\pmb{g}}_{A, ji}$ and elevation components $\hat{\pmb{g}}_{E, ji}$ of the estimated channel $\hat{\pmb{g}}_{ji} = \hat{\pmb{g}}_{A, ji} \otimes \hat{\pmb{g}}_{E,ji}$ are extracted by using the ``Nearest Kronecker product'' scheme \cite{Golub12Book, Loan92Book}. This scheme finds the solution of the minimization problem $\min_{\hat{\pmb{g}}_{A, ji}, \hat{\pmb{g}}_{E, ji}} \|\hat{\pmb{g}}_{ji} - \hat{\pmb{g}}_{A, ji} \otimes \hat{\pmb{g}}_{E, ji}\|_F$ based on the singular value decomposition. Accordingly, the azimuth component and the elevation component are computed as $\hat{\pmb{g}}_{A, ji} = \sqrt{\sigma_{ji, 1}}\pmb{u}_{ji,1}$ and $\hat{\pmb{g}}_{E, ji}= \sqrt{\sigma_{ji, 1}}\pmb{v}_{ji,1}$, respectively, where when defining the $N_{A, j}\times N_{E, j}$ matrix $\hat{\pmb{G}}_{ji} = \pmb{U}_{ji}\pmb{\Sigma}_{ji}\pmb{V}_{ji}$ whose elements are taken columnwise from $\hat{\pmb{g}}_{ji}$; $\sigma_{ji, 1}$ is the largest singular value component of the matrix $\pmb{\Sigma}_{ji}$; and $\pmb{u}_{ji,1}$ and $\pmb{v}_{ji,1}$ are corresponding left and right singular vectors of the matrix $\pmb{U}_{ji}$ and $\pmb{V}_{ji}$, respectively. With the estimated channel components $\hat{\pmb{g}}_{A, ji}$ and $\hat{\pmb{g}}_{E, ji}$, the effective power $\xi_{ji}^{(k)}(\pmb{\Omega}_{jk})$ in (\ref{eq:effch}) is decomposed into $\xi_{ji}^{(k)}(\pmb{\Omega}_{jk}) = (1-D_j)^2 \xi_{A, ji}^{(k)}(\pmb{\Omega}_{A, jk})\xi_{E, ji}^{(k)}(\pmb{\Omega}_{E, jk})$, where $\xi_{A, ji}^{(k)}(\pmb{\Omega}_{A, jk}) = \hat{\pmb{g}}_{A, ji}^H\pmb{\Omega}_{A, jk}\hat{\pmb{g}}_{A, ji}$ and $\xi_{E, ji}^{(k)}(\pmb{\Omega}_{E, jk}) = \hat{\pmb{g}}_{E, ji}^H\pmb{\Omega}_{E, jk}\hat{\pmb{g}}_{E, ji}$. 

At each outer $t$th iteration, two inner loops are employed in order to obtain the next iterate $\{\pmb{\Omega}_{ji}(t+1)\}$. The first is used to optimize the azimuth covariance matrices $\{\pmb{\Omega}_{A, ji}(t+1)\}$ for fixed elevation covariance matrices $\{\pmb{\Omega}_{E, ji}(t)\}$; while the second is used for optimizing the elevation covariance matrices $\{\pmb{\Omega}_{E, ji}(t+1)\}$ with fixed azimuth covariance matrices $\{\pmb{\Omega}_{A, ji}(t+1)\}$. For each inner loop, the SCA-based approach detailed in Algorithm \ref{al1} is applied to optimize the azimuth or elevation covariance matrices separately along with the auxiliary variables $(\pmb{\kappa}, \pmb{\chi})$.    

\section{Numerical Results}\label{sec:num}
\begin{figure}[t]
\begin{center}
\includegraphics[width=18cm]{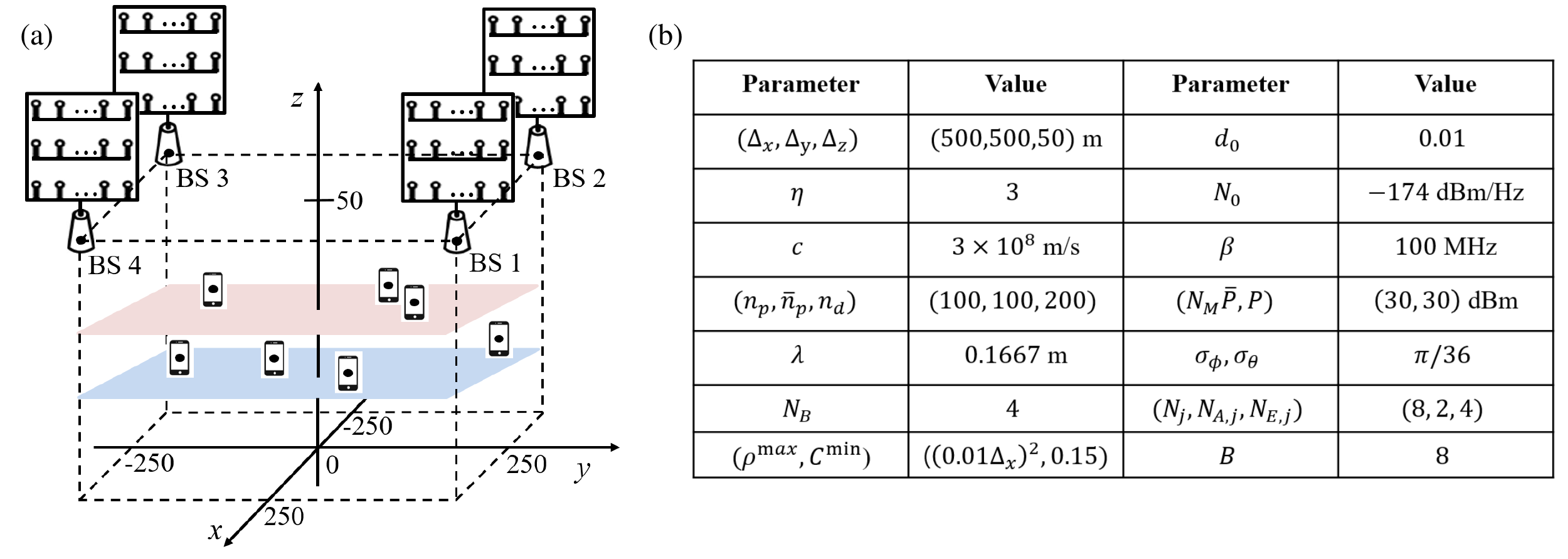}
\caption{(a) Set-up and (b) simulation parameters for the numerical results.} \label{fig:sim}
\end{center}
\end{figure}

In this section, we evaluate the performance of the unstructured and Kronecker beamforming design proposed in Section \ref{sec:beam_ideal} and Section \ref{sec:beam_kron}, respectively, in terms of the average sum-rate and the square root of average sum-SPE via Monte Carlo simulations. As illustrated in Fig. \ref{fig:sim}(a), we consider a network in which $N_B = 4$ BSs are placed at the vertices of square region of side length $\Delta_x = \Delta_y$ m at height $z_{B, j} = 50$ m for all BSs $j \in \mathcal{N}_B$, while the MSs are randomly uniformly distributed within a $\Delta_x \times \Delta_y \times \Delta_z$ cube. We consider both the maximization of the sum-rate under SPE constraints in problem {\bf{P1}} and the minimization of the sum-SPE under rate constraints in problem {\bf{P2}}. Furthermore, for reference, we include the upper bound of the solution of problem {\bf{P1}} obtained by removing the localization accuracy constraints (\ref{eq:P1_local}), as well as the lower bound on the optimal solution of problem {\bf{P2}} obtained by removing the data rate constraints (\ref{eq:P2_rate}). For both bounds, we consider the performance under unstructured beamforming. 

Unless stated otherwise, we consider the Kronecker channel model as described in Section \ref{sec:chk}. The parameters are summarized in Fig. \ref{fig:sim}(b) and described in this paragraph. We assume that each BS $j \in \mathcal{N}_B$ is equipped with $N_j = 8$ antennas consisting of $N_{A, j}=2$ horizontal antennas and $N_{E, j} = 4$ vertical antennas. We also set $\Delta_x = \Delta_y = 500$ m and $\Delta_z = 50$ m, and the reference distance $d_0$ in (\ref{eq:pl}) is $0.01$ so that the path loss at a distance of $100$ m is $\zeta = -60$ dB with path loss exponent $\eta = 3$. Moreover, we assume a noise level of $N_0 = -174$ dBm/Hz, the propagation speed $c = 3 \times 10^8$ m/s, an effective bandwidth of $\beta = 100$ MHz, and training and data phases with $\bar{n}_p = n_p = 100$ and $n_d = 200$ symbols. The wavelength is chosen to be $\lambda = 0.1667$ m, which corresponds to carrier frequency $1.8$ GHz, and the azimuth and elevation angular perturbation are set to $\sigma_\phi = \sigma_\theta = \pi/36$. The total uplink and downlink transmit powers are assumed to be $N_M\bar{P} = P = 30$ dBm and the number of output levels at the ADC/DACs at all the BSs are equal to $B = B_j=8$ for $j \in \mathcal{N}_B$. Identical requirements for localization accuracy $\rho^{\max} = \rho^{\max}_i = (0.01\Delta_x)^2$ in problem {\bf{P1}} and data rate $C^{\min} = C^{\min}_i = 0.15$ in problem {\bf{P2}} are applied to all MSs. 

\begin{figure*}[t]
\centering
\subfigure{\label{fig:B_Gch}\includegraphics[width=9cm]{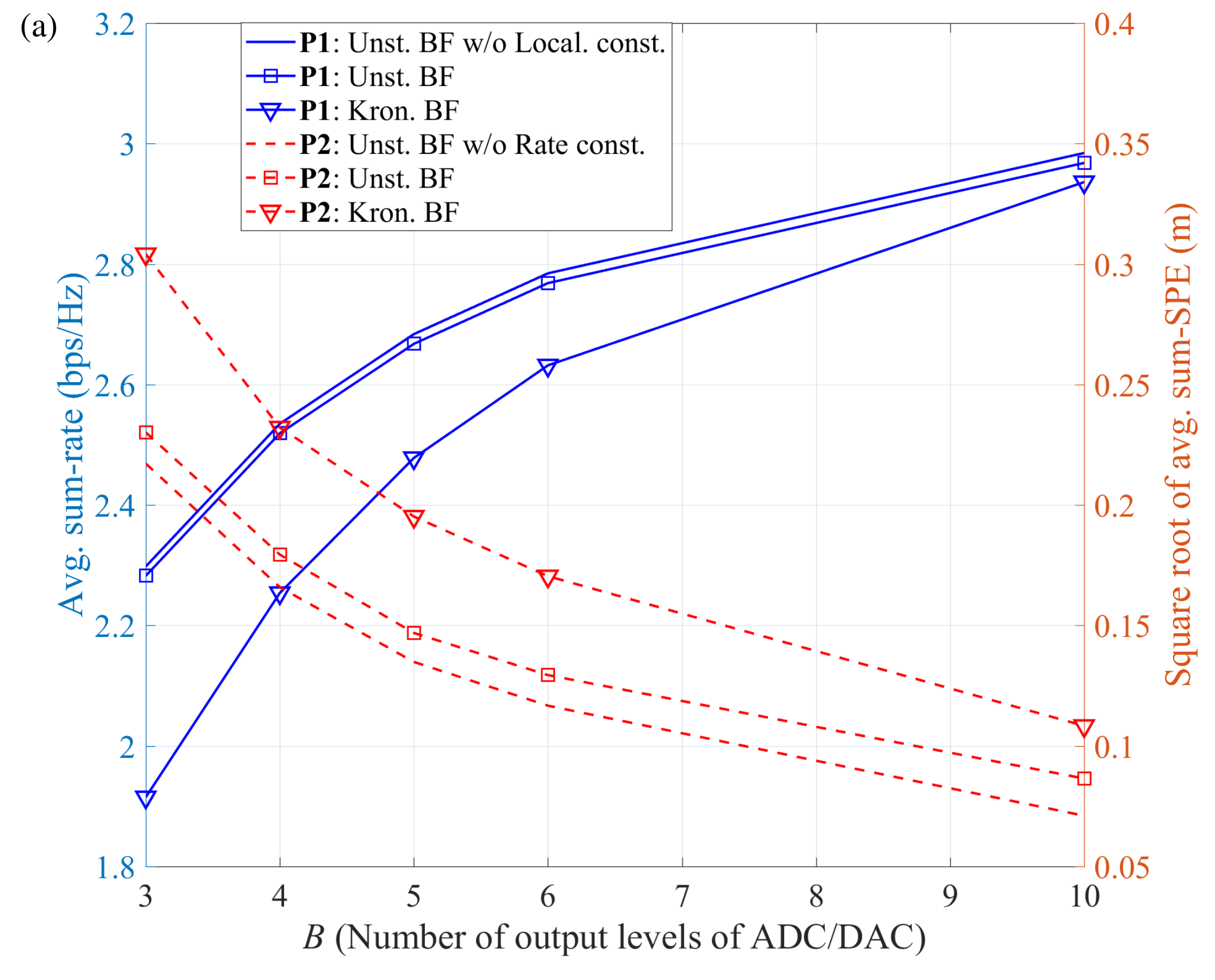}}
\subfigure{\label{fig:B_Kch}\includegraphics[width=9cm]{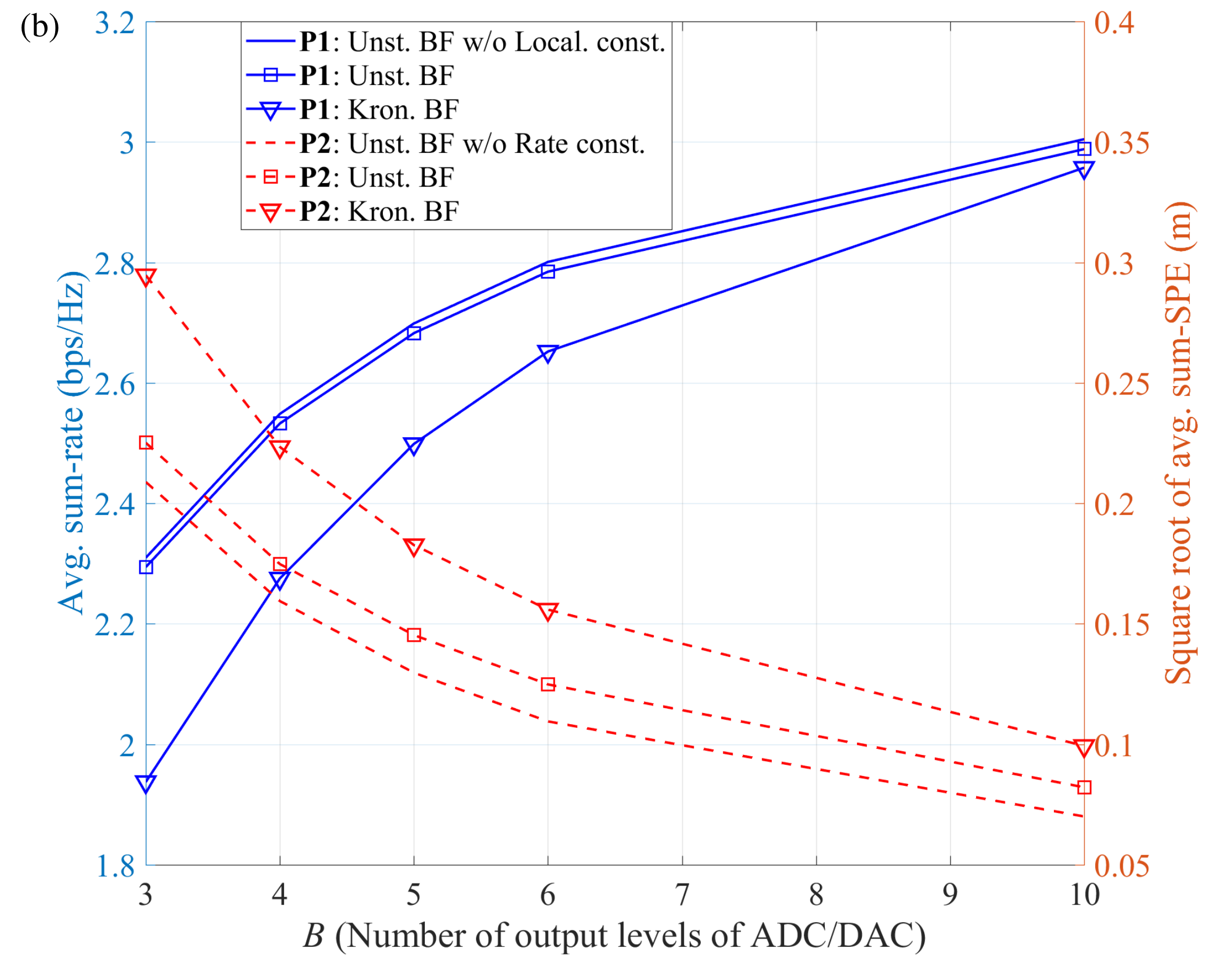}}\\
\caption{Average sum-rate for {\bf{P1}} and square root of average sum-SPE for {\bf{P2}} as a function of number $B$ of output levels of ADC/DACs at BSs with (a) correlated channel model and (b) Kronecker channel model ($N_M=2$ MSs randomly and uniformly distributed within the cube area).}
\label{fig:B}
\end{figure*}  

\begin{figure}[t]
\begin{center}
\includegraphics[width=10cm]{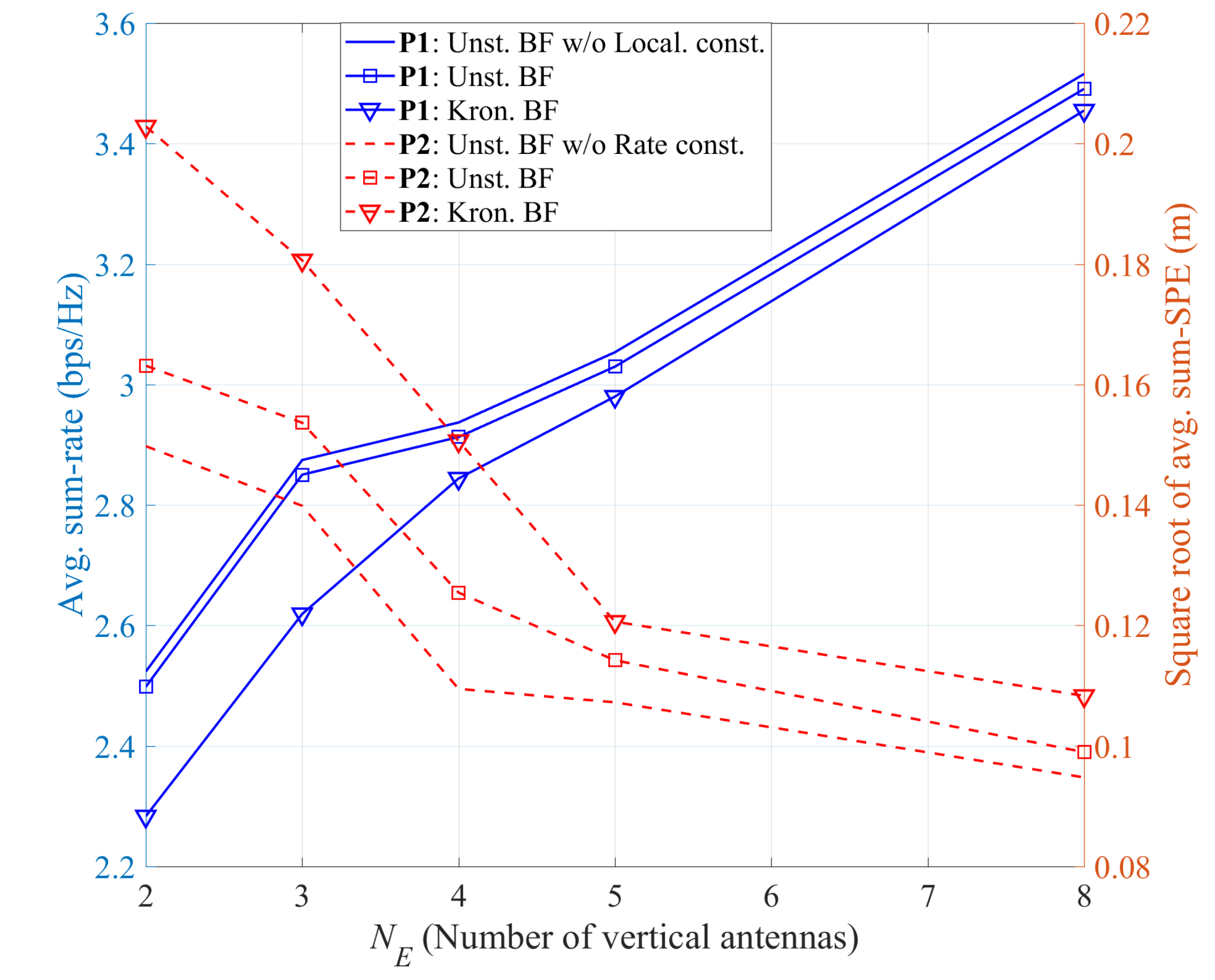}
\caption{Average sum-rate for {\bf{P1}} and square root of average sum-SPE for {\bf{P2}} as a function of number $N_E$ of vertical antennas at BSs (Kronecker channel model; $N_M=2$ MSs randomly and uniformly distributed within the cube area).} \label{fig:NE}
\end{center}
\end{figure}

\begin{figure}[t]
\begin{center}
\includegraphics[width=10cm]{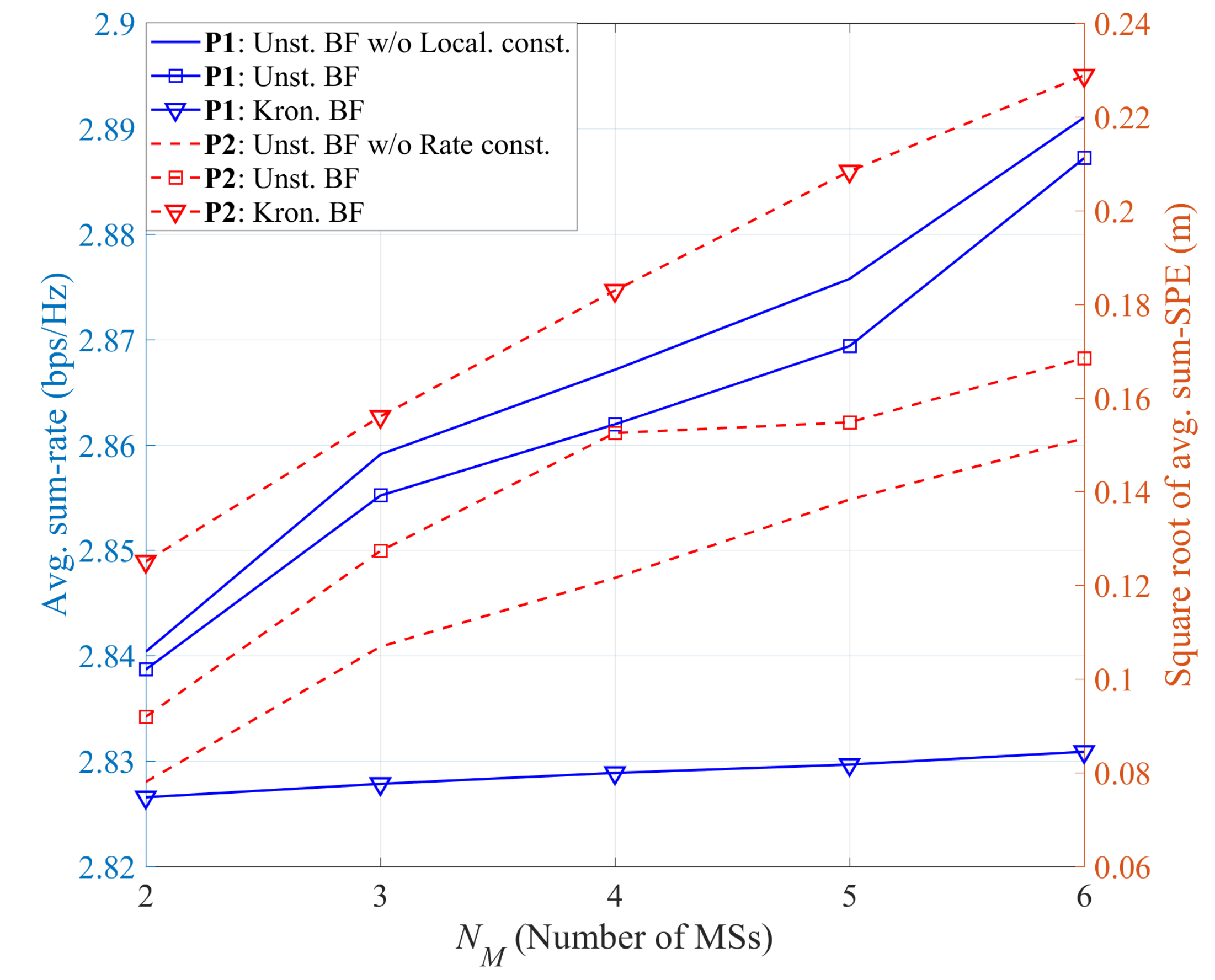}
\caption{Average sum-rate for {\bf{P1}} and square root of average sum-SPE for {\bf{P2}} as a function of number $N_M$ of MSs (Kronecker channel model; $N_M$ MSs randomly and uniformly distributed within the cube area).} \label{fig:NM}
\end{center}
\end{figure}

In Fig. \ref{fig:B}, we investigate the effect of the resolution $B$ of the ADC/DACs at the BSs with $N_M=2$ MSs in correlated channel model (Fig. \ref{fig:B_Gch}) and Kronecker channel model (Fig. \ref{fig:B_Kch}) by showing the rate and localization performance obtained from problems (\ref{eq:opt_rate}) and (\ref{eq:opt_local}), respectively. In general, as the resolution $B$ increases, both data rate and localization accuracy are improved. More interestingly, the proposed schemes with unstructured beamforming are able to accommodate localization or rate constraints with only minor performance degradations as compared to the reference case without such constraints. In contrast, reduced-complexity Kronecker solutions present a performance loss, which tends to decrease with $B$. This is also observed to be slightly increased in correlated channel model as compared to the Kronecker channel model. For instance, with $B=6$ bits, the performance loss of Kronecker beamforming over unstructured beamforming is $4.9$\% under the correlated channel model and $4.7$\% under the Kronecker channel model in terms of data rate optimization in {\bf{P1}}, while it amounts to $31.7$\% under the correlated channel model and $24.9$\% loss under the Kronecker channel model in terms of localization accuracy in {\bf{P2}}. 

Next, we consider the performance as a function of the number $N_E$ of vertical antennas in Fig. \ref{fig:NE} with the same setting with Fig. \ref{fig:B} but with $B = 8$ bits. Increasing the number $N_E$ of vertical antennas is seen to provide enhanced data rate and localization performance due to the larger available number of degrees of freedom. The figure confirms the main conclusions discussed above with regard to the effectiveness of the proposed optimal strategy. Furthermore, a larger $N_E$ is also observed to reduce the performance loss of Kronecker beamforming as compared to unstructured beamforming for rate optimization under localization constraints. This reduction is instead less noticeable for localization optimization under rate constraints. This can be explained by the fact that ensuring data rate requirements calls for the management of inter-MS interference, which is instead not an issue for positioning. In fact, localization accuracy is improved as long as the overall received power on all beams is increased. This suggests that guaranteeing minimal rate constraints requires a larger number of degrees of freedom than ensuring localization accuracy constraints.

Fig. \ref{fig:NM} investigates the same performance criteria as a function of the number $N_M$ of MSs. First, it is observed that a larger $N_M$ allows the achievement of an enhanced data rate performance, since the presence of more user offers an increased multiuser diversity that allows the transmission to MSs that have favorable channel conditions. In contrast, increasing $N_M$ yields a degraded performance in terms of localization accuracy. This is because each MS tends to receive less power for a larger $N_M$ due to the need to beamform to individual users in order to satisfy the rate constraints.  

\section{Concluding Remarks}\label{sec:con}
In this paper, we have investigated the problem of beamforming design for a TDD-based downlink system with FD-MIMO BSs equipped with limited-resolution front ends. Unlike existing works, we considered simultaneously the criteria of sum-rate data transmission and sum-SPE localization accuracy, under the assumption that localization is performed at the MSs based on downlink signals. Two types of beamforming designs are studied, namely \textit{unstructured beamforming} and a low-complexity \textit{Kronecker beamforming} solution, where the latter decomposes the beamforming vector for each BS into separate azimuth and elevation components. Extensive numerical results suggest that the proposed schemes can effectively ensure both data and localization performance criteria with a limited performance loss as compared to the case in which either data communication or positioning is considered. The interplay between rate and localization accuracy is seen to arise from their different requirements on beamforming design: While inter-MS interference management is required for the enhanced data rate in beamforming design, positioning only calls for the maximization of the power received at the users, irrespective of whether it is useful signal or interference. Optimized low-complexity Kronecker beamforming solutions are observed to achieve limited performance loss when the front-end resolution is large enough, the number of transmit antennas is large and the number of users is small. Among open issues left for future work, we mention the performance comparison of downlink positioning and uplink positioning in asynchronous scenarios.    
\appendices
\section{Derivations of $\gamma_1, \dots, \gamma_7$ in (\ref{eq:ch_R})}\label{app:ch}
In this appendix, by following \cite{Ying14ICC}, the definition of the $\gamma_1, \dots, \gamma_7$ is recalled for the channel model (\ref{eq:ch_R}). The small-scale fading channel vector $\pmb{h}_{ji}$ can be written as $\pmb{h}_{ji} = e^{j\varsigma_{ji}}\pmb{a}_{ji} \otimes \pmb{b}_{ji}$, where the phase shift $\varsigma_{ji}$ of transmit path between BS $j$ and MS $i$ is assumed to be uniformly distributed in $[0, 2\pi]$, and the steering vectors for y-axis and z-axis are given as $\pmb{a}_{ji} = [1 \,\,\, e^{-j\frac{2\pi d_A}{\lambda}\cos(\phi_{ji}+\Delta_{\phi_{ji}})\sin(\theta_{ji}+\Delta_{\theta_{ji}})} \cdots$ $e^{-j(N_{A, j}-1)\frac{2\pi d_A}{\lambda}\cos(\phi_{ji}+\Delta_{\phi_{ji}})\sin(\theta_{ji}+\Delta_{\theta_{ji}})}]^T$ and $\pmb{b}_{ji} = [1 \,\,\, e^{-j\frac{2\pi d_E}{\lambda}\cos(\theta_{ji}+\Delta_{\theta_{ji}})} \cdots e^{-j(N_{E, j}-1)\frac{2\pi d_E}{\lambda}\cos(\theta_{ji}+\Delta_{\theta_{ji}})}]^T$, respectively, for $j \in \mathcal{N}_B$ and $i \in \mathcal{N}_M$, with the azimuth angular perturbation $\Delta_{\phi_{ji}} \sim \mathcal{N}(0, \sigma_{\phi}^2)$ and elevation angular perturbation $\Delta_{\theta_{ji}} \sim \mathcal{N}(0, \sigma_{\theta}^2)$. As derived in \cite{Ying14ICC}, this yields the covariance matrix $\pmb{R}_{h_{ji}}$ in (\ref{eq:ch_R}), where $\gamma_1 = e^{j \frac{2\pi d_E}{\lambda}(p-k)\cos \theta_{ji}}e^{-\frac{1}{2}\left(\sigma_\theta\frac{2\pi d_E}{\lambda}\right)^2(p-k)^2\sin^2\theta_{ji}}$; $\gamma_2 = \frac{2\pi d_A}{\lambda}(q-l)\sin\theta_{ji}$; $\gamma_3 = \sigma_\theta\frac{2\pi d_A}{\lambda}(q-l)\cos\theta_{ji}$; $\gamma_4 = \frac{1}{2}\left(\sigma_\theta\frac{2\pi}{\lambda}\right)^2 d_Ed_A(p-k)(q-l)\sin(2\theta_{ji})$; $\gamma_5 = \gamma_3^2\sigma_\phi^2\sin^2\phi_{ji}+1$; $\gamma_6 = \gamma_4^2\sigma_\phi^2\sin^2\phi_{ji}+\cos\phi_{ji}$; $\gamma_7 = \gamma_3^2\cos^2\phi_{ji} - \gamma_4^2\sigma_\phi^2\sin^2\phi_{ji} -2\gamma_4\cos\phi_{ji}$.
\section{Derivations of (\ref{eq:crb})}\label{app:crb}
Here, we derive the performance metric (\ref{eq:crb}) for the localization accuracy of MS $i$. The MCRB is the trace of the inverse of the average EFIM, where the average, in this work, is with respect to the pilot signals \cite{Mengali94, Gini98}. To evaluate the average EFIM, we start by defining the unknown parameter vector for MS $i$ as $\pmb{\psi}_i=[\pmb{p}_{M,i}^T \,\,\, \pmb{\alpha}_{1i}^T \,\,\, \sigma^2_{\tilde{z}_{1i}}(\pmb{W}_1) \cdots \pmb{\alpha}_{N_Bi}^T \,\,\,$ $\sigma^2_{\tilde{z}_{N_Bi}}(\pmb{W}_{N_B})]^T$, where $\pmb{\alpha}_{ji}=[\Re\{\alpha^{(1)}_{ji}(\pmb{w}_{j1})\}$ $\,\, \Im\{\alpha^{(1)}_{ji}(\pmb{w}_{j1})\} \cdots \Re\{\alpha^{(N_M)}_{ji}(\pmb{w}_{jN_M})\} \,\, \Im\{\alpha^{(N_M)}_{ji}(\pmb{w}_{jN_M})\}]^T$. Computing the average EFIM requires to evaluate the quantity $\mathbb{E}_{\pmb{y}^p_{i}, \pmb{s}^p}[(\partial \ln f(\pmb{y}^p_{i} | \pmb{s}^p, \pmb{\psi}_{i})/\partial \pmb{\psi}_i)^2]$, which is averaged over the training sequences $\pmb{s}^p=\{\pmb{s}^p_{ji}\}_{j \in \mathcal{N}_B, i \in \mathcal{N}_M}$, with $\pmb{s}^p_{ji}$ being the vector representation of $s^p_{ji}(t) = \sum_{l=1}^{n_p}s^p_{ji}(l)p(t-lT_s)$. This is made difficult by the fact that the effective noise $\tilde{z}^p_{ji}(l)$ in (\ref{eq:y_dis_re}) is not independent of the useful signal and not Gaussian. 

To address this problem, we use the lower bound derived in \cite{Stein14SPL, Stein17TSP} that only requires the knowledge of the first and second moment of the system output. Accordingly, we have the inequality
\begin{subequations}\label{eq:FIM_lower}
\begin{eqnarray}
\mathbb{E}_{\pmb{y}^p_{i}, \pmb{s}^p}\left[\left(\frac{\partial \ln f\left(\pmb{y}^p_{i} \left| \pmb{s}^p, \pmb{\psi}_{i} \right.\right)}{\partial \pmb{\psi}_i}\right)^2\right] 
&=& \mathbb{E}_{\pmb{s}^p} \left[\mathbb{E}_{\pmb{y}^p_{i}|\pmb{s}^p} \left[ \left(\frac{\partial \ln f\left(\pmb{y}^p_{i} \left| \pmb{s}^p, \pmb{\psi}_{i} \right.\right)}{\partial \pmb{\psi}_i}\right)^2 \right]\right] \\
&\succeq& \mathbb{E}_{\pmb{s}^p} \left[ \left(\frac{ \partial\pmb{\mu}_{y^p_{i}}(\pmb{\psi}_i) }{\partial \pmb{\psi}_i} \right)^T \pmb{R}_{y^p_{i}}(\pmb{\psi}_i)^{-1} \left( \frac{ \partial\pmb{\mu}_{y^p_{i}}(\pmb{\psi}_i) }{\partial \pmb{\psi}_i} \right)\right] \triangleq \pmb{J}_{\pmb{\psi}_i}(\pmb{W}), \label{eq:ineq_FIM}
\end{eqnarray}
\end{subequations}
where the inequality relationship in (\ref{eq:ineq_FIM}) is obtained based on the Cauchy-Schwarz inequality; and $\pmb{\mu}_{y^p_{i}}(\pmb{\psi}_i)$ and $\pmb{R}_{y^p_{i}}(\pmb{\psi})$ are the first and second output moments given by $\pmb{\mu}_{y^p_{i}}(\pmb{\psi}_i) = \int \pmb{y}^p_{i}f\left(\pmb{y}^p_{i} \left| \pmb{s}^p, \pmb{\psi}_{i} \right.\right) d\pmb{y}^p_{i}$ and $\pmb{R}_{y^p_{i}}(\pmb{\psi}_i) = \int \left(\pmb{y}^p_{i}-\pmb{\mu}_{y^p_{i}}(\pmb{\psi}_i)\right)\left(\pmb{y}^p_{i}-\pmb{\mu}_{y^p_{i}}(\pmb{\psi}_i)\right)^T f\left(\pmb{y}^p_{i} \left| \pmb{s}^p, \pmb{\psi}_{i} \right.\right) d\pmb{y}^p_{i}$, respectively. Note that we have emphasized the dependence of (\ref{eq:ineq_FIM}) on the beamforming vectors $\pmb{W}=\{\pmb{W}_j\}_{j \in \mathcal{N}_B}$. To evaluate (\ref{eq:FIM_lower}), we relate the parameter vector $\pmb{\psi}_i$ to the larger parameter vector $\tilde{\pmb{\psi}}_i$, where $\tilde{\pmb{\psi}}_i = [\tilde{\pmb{\psi}}_{1i}^T \cdots \tilde{\pmb{\psi}}_{N_Bi}^T]^T$ with $\tilde{\pmb{\psi}}_{ji}=[\tau_{ji} \,\,\pmb{\alpha}_{ji}^T \,\, \sigma_{\tilde{z}_{ji}}^2(\pmb{W}_j)]^T$. When the MS is localizable, this mapping is a bijection \cite{JSA14TVT, JSA15TVT, JSA16IET}, and this reparameterization allows us to rewrite the matrix $\pmb{J}_{\pmb{\psi}_i}(\pmb{W})$ in (\ref{eq:FIM_lower}) as $\pmb{J}_{\pmb{\psi}_i}(\pmb{W}) = \pmb{T}\pmb{J}_{\tilde{\pmb{\psi}}_i}(\pmb{W})\pmb{T}^T$, with $\pmb{T}$ being the Jacobian matrix for the transformation from $\pmb{\psi}_i$ to $\tilde{\pmb{\psi}}_i$ and $\pmb{J}_{\tilde{\pmb{\psi}}_i}(\pmb{W})$ being defined in the same way of $\pmb{J}_{\pmb{\psi}_i}(\pmb{W})$ in (\ref{eq:ineq_FIM}) with $\tilde{\pmb{\psi}}_i$ in lieu of $\pmb{\psi}_i$. By applying the standard Schur complement condition \cite{Zhang05Book} to the matrix $\pmb{J}_{\pmb{\psi}_i}(\pmb{W})$, we obtain the following lower bound on the submatrix of $\pmb{J}_{\pmb{\psi}_i}(\pmb{W})$ in (\ref{eq:efim}) corresponding to the MS $i$'s position $\pmb{p}_{M, i}$. The derivation of (\ref{eq:efim}) from (\ref{eq:FIM_lower}) follows using the same steps as in \cite{JSA14TVT, JSA15TVT, JSA16IET}, which yields the performance metric for the SPE in (\ref{eq:spe}) of MS $i$ as the MCRB in (\ref{eq:crb}).
\section{Derivation of Problem {\bf{P1-2}}}\label{app:P1}
In this appendix, we derive the SCA algorithm in Algorithm \ref{al1} for problem {\bf{P1-1}}. We first observe that the objective function in (\ref{eq:R_obj1}) is the sum of rates $C_{ji}(\pmb{v})$, each having the DC form 
\begin{equation}\label{eq:R_DC}
C_{ji}(\pmb{v}) = \frac{n_d}{N_Bn} \left( f^{+}_{ji}(\pmb{v}) - f^{-}_{ji}(\pmb{v})\right),
\end{equation}
where
\begin{subequations}\label{eq:R_DC_func}
\begin{eqnarray}
&& \hspace{-1.6cm} f^{+}_{ji}(\pmb{v}) = \log_2\left( \sum_{k \in \mathcal{N}_M} \xi^{(k)}_{ji}(\pmb{\Omega}_{jk}) +   N_0 + \sum_{k \in \mathcal{N}_M}(1-D_j)^2\text{tr}\{\pmb{R}_{\Delta_{g_{ji}}}\pmb{\Omega}_{jk}\} + \sum_{n=1}^{N_j} \kappa_{jn}\text{tr}\left\{\pmb{E}_n\pmb{R}_{g_{ji}}\pmb{E}_n^T\right\} \right) \\
&& \hspace{-1.6cm} \text{and} \nonumber\\
&&\hspace{-1.6cm} f^{-}_{ ji}(\pmb{v}) = \log_2\left( \sum_{k \in \mathcal{N}_M, k \neq, i} \xi^{(k)}_{ji}(\pmb{\Omega}_{jk}) + N_0 + \sum_{k \in \mathcal{N}_M}(1-D_j)^2\text{tr}\{\pmb{R}_{\Delta_{g_{ji}}}\pmb{\Omega}_{jk}\} + \sum_{n=1}^{N_j} \kappa_{jn}\text{tr}\left\{\pmb{E}_n\pmb{R}_{g_{ji}}\pmb{E}_n^T\right\} \right).
\end{eqnarray} 
\end{subequations}
Then, given $\pmb{v}(t)$ for the $t$th feasible iterate, we obtain the strongly concave surrogate function $\hat{C}(\pmb{v}; \pmb{v}(t))$ in (\ref{eq:hatC}) by applying Lemma 1. 

For the non-convex constraint (\ref{eq:R_chi1}), we first define the function 
\begin{subequations}\label{eq:EFIM_chi}
\begin{eqnarray}
g_{ji}(v_1, \pmb{v}_2) &\triangleq& \chi_{ji}\left(\sum_{k \in \mathcal{N}_M}(1-D_j)^2\text{tr}\{\pmb{R}_{\Delta_{g_{ji}}}\pmb{\Omega}_{jk}\}  +  \sum_{n=1}^{N_j} \kappa_{jn}\text{tr}\left\{\pmb{E}_n\pmb{R}_{g_{ji}}\pmb{E}_n^T\right\}\right) \\
& = & \frac{1}{2}\left( h_{1, ji}(v_1) + h_{2, ji}(\pmb{v}_2) \right)^2 - \frac{1}{2}\left( h_{1, ji}^2(v_1) + h_{2, ji}^2(\pmb{v}_2) \right),
\end{eqnarray}
\end{subequations}
for $j \in \mathcal{N}_B$ and $i \in \mathcal{N}_M$, where $h_{1, ji}(v_1) = \chi_{ji}$; $h_{2, ji}(\pmb{v}_2) = \sum_{k \in \mathcal{N}_M}(1-D_j)^2\text{tr}\{\pmb{R}_{\Delta_{g_{ji}}}\pmb{\Omega}_{jk}\}  +  \sum_{n=1}^{N_j} \kappa_{jn}$ $\text{tr}\left\{\pmb{E}_n\pmb{R}_{g_{ji}}\pmb{E}_n^T\right\}$; $v_1 = \chi_{ji}$ and $\pmb{v}_2 = \{\pmb{\Omega}_{j1}, \dots, \pmb{\Omega}_{jN_M}, \kappa_{j1}, \dots, \kappa_{jN_j}\}$. Given a feasible solution $\pmb{v}(t)$, using Lemma 2, we then derive the upper bound $\bar{g}_{ji}(v_1, \pmb{v}_2;$ $v_1(t), \pmb{v}_2(t)) \ge g_{ji}(v_1, \pmb{v}_2)$, where the left-hand side is defined as (\ref{eq:EFIM_chi_upper}). Consequently, since the constraint (\ref{eq:R_chi1}) can be written as $g_{ji}(v_1, \pmb{v}_2) \le \sum_{k \in \mathcal{N}_M} \xi^{(k)}_{ji}(\pmb{\Omega}_{jk}) - \chi_{ji}N_0$, it can be convexified as $\bar{g}_{ji}(v_1, \pmb{v}_2; v_1(t), \pmb{v}_2(t)) \le \sum_{k \in \mathcal{N}_M}\xi^{(k)}_{ji}(\pmb{\Omega}_{jk}) - \chi_{ji}N_0$ in (\ref{eq:R_chi2}) while satisfying the conditions of the SCA method (Lemma 2). 
\section{Derivation of Problem {\bf{P2-2}}}\label{app:P2}
In this appendix, we provide technical details on the derivations of the SCA method in Algorithm \ref{al1} for the sum-SPE minimization problem. We start by relaxing the rank constraint on $\pmb{\Omega}$ and obtain the equivalent problem:  
\begin{subequations}\label{eq:opt_local1}
\begin{eqnarray}
&& \hspace{-0.7cm} \text{(\bf{P2-1}):} \hspace{0.2cm} \underset{\pmb{\Omega}, \pmb{\kappa}, \pmb{\chi}}{\text{minimize}} \hspace{0.2cm} \sum_{i \in \mathcal{N}_M}\frac{1}{N_s}\sum_{m=1}^{N_s}\text{tr}\left\{\left(\frac{8 \pi^2 n_p \beta^2}{c^2} \sum_{j=1}^{N_B} \pmb{J}_{\phi, \theta} (\phi_{ji,m}, \theta_{ji,m}) \chi_{ji} \right)^{-1}\right\} \label{eq:L_obj1}\\
&& \hspace{-0.7cm}  \text{s.t.} \hspace{0.2cm} \frac{n_d}{N_Bn} \sum_{j \in \mathcal{N}_B} \log_2\left( \sum_{k \in \mathcal{N}_M} \xi^{(k)}_{ji}(\pmb{\Omega}_{jk}) + N_0 + \sum_{k \in \mathcal{N}_M}(1-D_j)^2\text{tr}\{\pmb{R}_{\Delta_{g_{ji}}}\pmb{\Omega}_{jk}\} +  \sum_{n=1}^{N_j} \kappa_{jn}\text{tr}\left\{\pmb{E}_n\pmb{R}_{g_{ji}}\pmb{E}_n^T\right\} \right) \nonumber\\
&& \hspace{-0.7cm}  - \log_2\left( \sum_{k \in \mathcal{N}_M, k \neq, i} \xi^{(k)}_{ji}(\pmb{\Omega}_{jk}) +   N_0 + \sum_{k \in \mathcal{N}_M}(1-D_j)^2\text{tr}\{\pmb{R}_{\Delta_{g_{ji}}}\pmb{\Omega}_{jk}\} + \sum_{n=1}^{N_j} \kappa_{jn}\text{tr}\left\{\pmb{E}_n\pmb{R}_{g_{ji}}\pmb{E}_n^T\right\} \right) \ge C^{\min}_i, \hspace{0.2cm} i \in \mathcal{N}_M \label{eq:L_rate1}\nonumber\\
&&\\
&& \hspace{-0.7cm} \text{(\ref{eq:R_pt1}) - (\ref{eq:R_chi_plus1}).} 
\end{eqnarray}
\end{subequations}  
The problem {\bf{P2-1}} is not convex due to the non-convex constraints (\ref{eq:L_rate1}) and (\ref{eq:R_chi1}). In order to apply the SCA method, we first recall the following lemma. 

\textit{Lemma 3} ([43, Example 3]): Consider a non-convex constraint $g_i(\pmb{x}) \ge 0$, where $g_i(\pmb{x})$ has the DC structure, namely  $g_i(\pmb{x}) = g^{+}_i(\pmb{x}) - g^{-}_i(\pmb{x})$, with $g^{+}_ i(\pmb{x})$ and $g^{-}_i(\pmb{x})$ being concave and continuously differentiable. For any $\pmb{y}$ in the domain of $g_i(\pmb{x})$, a concave lower approximation $\bar{g}_i(\pmb{x}; \pmb{y}) \le g_i(\pmb{x})$ that guarantees the requirements \cite[Assumption 3]{Scutari14Arxiv} of the SCA algorithm is given as 
\begin{equation}\label{eq:SCA_con2}
\bar{g}_i(\pmb{x}; \pmb{y}) \triangleq g^{+}_i(\pmb{x}) - g^{-}_i(\pmb{y}) - \nabla_{\pmb{x}}g^{-}_i(\pmb{y})^T(\pmb{x}-\pmb{y}).  
\end{equation}  

We define the set of optimization variables for problem {\bf{P2-1}} as $\pmb{v} = (\pmb{\Omega}, \pmb{\kappa}, \pmb{\chi})$ and $\pmb{v}(t) = (\pmb{\Omega}(t), \pmb{\kappa}(t), \pmb{\chi}(t))$ for the $t$th iterate within the feasible set of the problem {\bf{P2-1}}. In addition, the non-convex constraints (\ref{eq:L_rate1}) can be written as $\sum_{j \in \mathcal{N}_B} C_{ji}(\pmb{v}) \ge C^{\min}_i$ in (\ref{eq:L_rate1}), where $C_{ji}(\pmb{v})$ is a DC function with $f^{+}_{ji}(\pmb{v})$ and $f^{-}_{ji}(\pmb{v})$ in (\ref{eq:R_DC}) as $g^{+}_{ji}(\pmb{v})$ and $g^{-}_{ji}(\pmb{v})$, respectively. As a result, using Lemma 3, a concave lower bound $\bar{C}_{ji}(\pmb{v}; \pmb{v}(t)) \le C_{ji}(\pmb{v})$ can be derived for use in the SCA algorithm as     
\begin{equation}\label{eq:barC} 
\bar{C}_{ji}(\pmb{v}; \pmb{v}(t)) \triangleq  \frac{n_d}{N_Bn} \left( g^{+}_{ji}(\pmb{v}) - g^{-}_{ji}(\pmb{v}(t)) - \nabla_{\pmb{v}} g^{-}_{ji}(\pmb{v}(t))^T(\pmb{v} - \pmb{v}(t)) \right),
\end{equation}
where $\nabla_{\pmb{v}} g^{-}_{ji}(\pmb{v}(t))^T(\pmb{v} - \pmb{v}(t))$ is calculated as (\ref{eq:fm_der}). For the non-convex constraints (\ref{eq:R_chi1}), a convex upper bound $\bar{g}_{ji}(v_1, \pmb{v}_2; v_1(t), \pmb{v}_2(t)) \ge g_{ji}(v_1, \pmb{v}_2)$ in (\ref{eq:EFIM_chi_upper}) is obtained using Lemma 2, yielding the convex constraint (\ref{eq:R_chi2}). Consequently, for a feasible $\pmb{v}(t)$, we obtain problem {\bf{P2-2}} in (\ref{eq:opt_local2}).

\bibliographystyle{IEEEtran}
\bibliography{JSAref}

\end{document}